\documentclass[%
 reprint,
 amsmath,amssymb,
 aps,
 showkeys, 
]{revtex4-2}

\usepackage{graphicx}
\usepackage{dcolumn}
\usepackage{bm}

\newcommand{\subcaption}[1]{%
  \vspace{4pt}
  \parbox{\linewidth}{ #1}%
}

\usepackage[colorlinks,
            linkcolor=red,
            anchorcolor=blue,
            citecolor=green
            ]{hyperref}

\begin{document}

\preprint{APS/123-QED}




\title{Towards Realistic Detection Pipelines of Taiji: New Challenges in Data Analysis and High-Fidelity Simulations of {Space-Based} Gravitational Wave Antenna}







\author{Minghui Du$^1$}
\email{duminghui@imech.ac.cn}%
\author{Pengcheng Wang$^{2,3}$}
\author{Ziren Luo$^{1,4,5}$}
\author{Wen-Biao Han$^{4,6,7,8}$}
\author{Xin Zhang$^{9,10,11}$}
\author{Xian Chen$^{12,13}$}
\author{Zhoujian Cao$^{8,14,15}$}
\author{Yonghe Zhang$^2$}
\author{He Wang$^{4,16}$}
\author{Xiaodong Peng$^{8,17}$}
\author{Li-E Qiang$^{17}$}
\author{Ke An$^{2,3}$}
\author{Yidi Fan$^{2,3}$}
\author{Jiafeng Zhang$^{17}$}
\author{Liang-Gui Zhu$^{13}$}
\author{Ping Shen$^{6,8}$}
\author{Qianyun Yun$^{6,18,19}$}
\author{Xiao-Bo Zou$^8$}
\author{Ye Jiang$^{6,7}$}
\author{Tianyu Zhao$^1$}
\author{Yong Yuan$^1$}
\author{Xiaotong Wei$^1$}
\author{Yuxiang Xu$^1$}
\author{Bo Liang$^1$}
\author{Peng Xu$^{1,4,5,20}$}
\email{xupeng@imech.ac.cn}
\author{Yueliang Wu$^{16,18,21}$}
\email{ylwu@itp.ac.cn}

\affiliation{$^1$Center for Gravitational Wave Experiment, National Microgravity Laboratory, Institute of Mechanics, Chinese Academy of Sciences, Beijing 100190, China}
\affiliation{$^2$Innovation Academy for Microsatellites of Chinese Academy of Sciences, Shanghai, 201304, China}
\affiliation{$^3$Key Laboratory for Satellite Digitalization Technology, Chinese Academy of Sciences, Shanghai, 201210, China}
\affiliation{$^4$Taiji Laboratory for Gravitational Wave Universe (Beijing/Hangzhou), University of Chinese Academy of Sciences (UCAS), Beijing 100049, China}
\affiliation{$^5$Key Laboratory of Gravitational Wave Precision Measurement of Zhejiang Province, Hangzhou Institute for Advanced Study, UCAS, Hangzhou,  310024, China}
\affiliation{$^6$Shanghai Astronomical Observatory, Shanghai 200030, China}
\affiliation{$^7$School of Astronomy and Space Science, University of Chinese Academy of Sciences, Beijing 100049, China}
\affiliation{$^8$School of Fundamental Physics and Mathematical Sciences, Hangzhou Institute for Advanced Study, UCAS, Hangzhou 310024, China}
\affiliation{$^9$Key Laboratory of Cosmology and Astrophysics (Liaoning) \& College of Sciences, Northeastern
University, Shenyang 110819, China}
\affiliation{$^{10}$National Frontiers Science Center for Industrial Intelligence and Systems Optimization, Northeastern University, Shenyang 110819, China}
\affiliation{$^{11}$Key Laboratory of Data Analytics and Optimization for Smart Industry (Ministry of Education),
Northeastern University, Shenyang 110819, China}
\affiliation{$^{12}$Astronomy Department, School of Physics, Peking University, 100871 Beijing, People's Republic of China}
\affiliation{$^{13}$Kavli Institute for Astronomy and Astrophysics at Peking University, 100871 Beijing, People's Republic of China}
\affiliation{$^{14}$School of Physics and Astronomy, Beijing Normal University, Beijing, 100875, China}
\affiliation{$^{15}$Institute for Frontiers in Astronomy and Astrophysics,
Beijing Normal University, Beijing, 102206, China}
\affiliation{$^{16}$International Centre for Theoretical Physics Asia-Pacific (ICTP-AP), University of Chinese Academy of Sciences (UCAS), Beijing 100049, China}
\affiliation{$^{17}$Key Laboratory of Electronics and Information Technology for Space System, National Space Science Center, Chinese Academy of Sciences, Beijing 100190, China}
\affiliation{$^{18}$Hangzhou Institute for Advanced Study, University of Chinese Academy of Sciences, Hangzhou 310124, China}
\affiliation{$^{19}$School of Physics and Astronomy, Shanghai Jiao Tong University, Shanghai 200240, China}
\affiliation{$^{20}$Lanzhou Center of Theoretical Physics, Lanzhou University, Lanzhou 730000, China}
\affiliation{$^{21}$Institute of Theoretical Physics, Chinese Academy of Sciences, Beijing 100190, China}

\date{\today}

\begin{abstract}
{Taiji, a Chinese space-based gravitational wave (GW) detection project, aims to explore the millihertz GW universe with unprecedented sensitivity.
By observing astrophysical and cosmological sources including Galactic binaries, massive black hole binaries, extreme mass-ratio inspirals, and stochastic gravitational wave backgrounds, \emph{etc}, Taiji is expected to deliver transformative insights into astrophysics, cosmology, and fundamental physics. 
However, Taiji’s data analysis faces unique challenges compared to ground-based detectors like LIGO-Virgo-KAGRA, such as the overlap of numerous signals, extended data durations, more rigorous accuracy requirements for the waveform templates, incompletely characterized noise spectra, non-stationary noises, and various data anomalies. 
Taking Taiji as a representative example, 
this paper reviews the data characteristics and data analysis challenges of space-based GW detection, and introduces the second round of Taiji Data Challenge, a collection of simulation datasets designed as a shared platform for resolving these critical issues.}
This platform distinguishes from previous works by the systematic integration of  orbital  dynamics based on a full drag-free and attitude control simulation, extended noise sources, more complicated and overlapping GW signals, second-generation time-delay interferometry, and the coupling effect of time-varying armlengths, \emph{etc}.
Concurrently released is the open-source toolkit \texttt{Triangle} (available at \href{https://github.com/TriangleDataCenter}{\textcolor{blue}{https://github.com/TriangleDataCenter}}), which offers  the  capabilities for customized simulation of signals, noises and other instrumental effects. 
By taking a step further towards realistic detection,  Taiji Data Challenge II and \texttt{Triangle} altogether serve as a new testbed,  supporting  the development of   Taiji's global analysis and end-to-end pipelines, and ultimately  bridging the gaps between observation and scientific objectives.
\end{abstract}

\maketitle


\section{Introduction}\label{sec:1}

{In the last decade, the LIGO-Virgo-KAGRA (LVK) ground-based gravitational wave (GW) detection network~\cite{Aasi_2015_aligo,Acernese_2015_avirgo,Akutsu2021KAGRA} has opened a new era of astrophysics with the  detection of over  a hundred GW events in the the Hz - kHz frequency band, originating from stellar-mass  binary black holes, binary neutron stars, and black hole-neutron stars.
More recently, in 2023,  multiple international pulsar timing array (PTA) collaborations  announced key evidence for nHz stochastic GW background (SGWB)~\cite{pta_nanograv,pta_epta,pta_ppta,pta_cpta}, marking a major breakthrough in the detection of low-frequency GWs. 
Between these frequency regimes lies the mHz band, which is rich in both astrophysical and cosmological sources, and  accessing this band is crucial for completing our understanding of the GW universe. 
Several ongoing space-based GW detection projects are designed to target the mHz band, including 
the Laser Interferometer Space Antenna (LISA)~\cite{AmaroSeoane2017LISA,Baker:2019nia},  Taiji~\cite{taiji_0,taiji_1,taiji_2},  and {TianQin}~\cite{TianQin1}, \emph{etc}~\cite{chinese_spacebased_gw_detection,lisa_taiji_tianqin_network1}.}


LISA is currently the most established international space-based GW detection mission, which has already been approved by the European Space Agency to enter the engineering phase, preparing for the launch in 2035~\cite{LISA_gets_go}.
{Meanwhile, the Taiji project, initiated in 2008 by the Chinese Academy of Sciences (CAS), is also stated for launch in the 2030s~\cite{taiji_0,taiji_2}.}
The baseline design of Taiji comprises three spacecrafts (SCs) forming a triangular constellation with 3-million-kilometer arms.
The constellation operates in a heliocentric orbit and leads the Earth by approximately $20^\circ$. 
As a recent milestone, Taiji-1, a technology demonstration satellite of Taiji~\cite{taiji1}, was launched in 2019 and had completed its mission to verify Taiji's key payloads and technologies, including laser interferometor~\cite{taiji1_laser_interferometry}, inertial sensor~\cite{taiji1_inertial_sensor}, micro thruster~\cite{taiji1_micro_thruster}, drag-free control~\cite{taiji1_drag_free}, and data processing~\cite{taiji1_data,taiji1_grs_calibration}, \emph{etc}.
Moreover, during its free fall, Taiji-1 also operated as a gravity recovery satellite and  produced the global gravity model TJGM-r1911~\cite{taiji1_gravfield}.
All these advancements  pave the paths towards the final Taiji mission.

The scientific operation of Taiji will last for at least 5 years, during which it will be observing burst, continuous, and stochastic GW signals in the 0.1 mHz - 1 Hz band, originating from various  sources, including  $\mathcal{O}(10^7)$  extra-Galactic and Galactic binaries (GBs) ($\mathcal{O}(10^4)$  resolvable, others forming a confusion foreground)~\cite{taiji_confusion_noise}, {$\mathcal{O}(10)$ - $\mathcal{O}(10^2)$ massive black hole binaries (MBHBs)~\cite{TaijiStandardSiren,TaijiLISAStandardSiren}}, {$\mathcal{O}(1)$ - $\mathcal{O}(10^3)$ extreme mass-ratio inspirals (EMRIs, the number is inferred from the research on LISA~\cite{emri_rate})}, {the inspiral phase of $\mathcal{O}(1)$ - $\mathcal{O}(10)$ stellar-mass black hole binaries (sBHBs)~\cite{sbhb_rate1}}, 
{as well as the astrophysical and/or cosmological  SGWBs~\cite{PhysRevD.109.063520,Chen:2024jca}, \emph{etc}}.

{Unlike the ``noise-dominated'' regime of LVK observations, the data of space-based detectors Taiji, LISA and TianQin are anticipated as ``signal-dominated'', which exhibit  several distinctive  features.} 
Firstly, as the signals overlap in both time and frequency domains, source-by-source analysis is generally impractical, making it inevitable to conduct joint analysis of multiple sources~\cite{Littenberg_prototype_2023}.
Secondly, a considerable fraction of signals have  high signal-to-noise ratios (SNRs) up to $\mathcal{O}(10^2)$ - $\mathcal{O}(10^3)$, 
thereby necessitating rigorous modeling of waveforms and detector responses~\cite{LISA_waveform_whitepaper}.
Thirdly, the majority of  detectable signals  are continuous GWs with observation timescales of months to years (\emph{e.g.} EMRIs, GBs), during which  glitches, data gaps, and non-stationary noises are anticipated to frequently occur~\cite{Castelli:2024sdb}.  
Fourthly, the astrophysical confusion foreground and instrumental noises in space environments remain incompletely understood, and during the future in-orbit operational phase, we might not have the luxury to monitor and characterize noise properties with extra auxiliary methods~\cite{unknown_noise}. 
Lastly, prior to scientific data analysis, data pre-processing constitutes a crucial stage for space-based GW detections, which  undertakes  the tasks of suppressing primary noises, calibrating key operational parameters, and assessing operational status and data quality, \emph{etc}, with  several of them  inherently  coupled with scientific analysis {(see \emph{e.g.} Ref.~\cite{Littenberg_prototype_2023})}. 
Nevertheless, a comprehensive end-to-end data analysis pipeline from raw data to scientific outcomes remains unestablished. 
As will be detailed in Section~\ref{sec:3}, 
these features  pose significant challenges unresolved by the conventional  methodologies used in ground-based detections, 
and  have been  recognized internationally as critical  problems in the field. 
The urgency for solving these  challenges further intensify with LISA's transition to engineering phase and  China's Taiji and {TianQin} missions pending approval.  

Facing the limitations of traditional data analysis methods in disentangling overlapping signals and handling  complex noises and anomalies, 
``Data Challenge'' serves as a proactive strategy to discover and address the issues in advance. 
Based on the collaborative  efforts of Taiji simulation group, by creating mock datasets  that encapsulate the complexity of Taiji's measurements, Taiji Data Challenge (TDC) aims to  establish a standardized platform for the Taiji Scientific Collaboration and broader researchers to develop specialized algorithms and validate data analysis tools.
TDC is expected to drive advancements in techniques and methodologies including time-delay interferometry (TDI), Bayesian parameter estimation,  machine learning, \emph{etc},
hence playing  a critical role in achieving  Taiji’s scientific objectives to reveal cosmic history, probe astrophysical phenomena,  test  fundamental theories, and  unlock transformative insights into the dark universe.

In recent years, exploration {has been made}   along this path, with well-known examples being the simulation datasets: mock LISA data challenge (MLDC)~\cite{MLDC}, LISA data challenge (LDC)~\cite{LDC},  the first round of Taiji data challenge (TDC I)~\cite{TDCI}, as well as the multi-mission science data simulator for space-based GW detection \texttt{GWSpace}~\cite{Li_GWSpace_2023}.
With the publication of ``global analyses'' on LDC2a  
by multiple groups~\cite{Littenberg_prototype_2023,Strub_global_2024,Katz_efficient_2024,Deng:2025wgk}, LDC has basically fulfilled its purpose as a simulation dataset based on simplified and idealized orbit configurations, instrumental noises and GW waveforms. 
In view of the continuous development of space-based GW detection missions, 
and  the high-complexity of detector system  that critically impacts  GW detection and data analysis, 
we believe it is now necessary to take a further step to introduce more realities and complexities into the mock data of Taiji.  
Consequently,  the  second round of Taiji data challenge (TDC II) is released~\cite{TDCII_website}, aiming  to explore and address the ``new'' challenges in the upcoming stage. 
Alongside mock datasets, an open-source toolkit ``\texttt{Triangle}'' used in the creation of TDC II will also be made public.  
\texttt{Triangle} is a prototype simulator for the data of space-based GW detectors.
It offers visibility  into the models behind TDC II, and enables researchers to simulate a wide range of scenarios that go beyond the default  settings of TDC II, hence supporting broader  explorations on data analysis and GW sciences.  
Altogether, this new simulation testbed is anticipated to assist in  bridging  the gaps between Taiji's realistic observations and scientific objectives. 

This paper is structured as follows: 
Section~\ref{sec:2} outlines Taiji's mission design and scientific data flow. 
{Taking Taiji as a representative example, Section~\ref{sec:3} systematically reviews the  key challenges for the data analysis of space-based GW detection.}
Motivated by these challenges, TDC II is introduced in Section~\ref{sec:4}, which includes 5 groups of datasets, each {targeting}  specific topics listed in Section~\ref{sec:3}. 
The models and simulation framework  used in mock data generation are also provided.
Section~\ref{sec:5} describes the open-source \texttt{Triangle} toolkit. 
The concluding remarks and outlooks for future development are presented in Section~\ref{sec:6}.
The Appendix illustrate a simple  parameter estimation task on TDC II ``verification'' dataset  using the tools provided in  \texttt{Triangle}.

\begin{figure}
    \centering
    \includegraphics[width=0.8\linewidth]{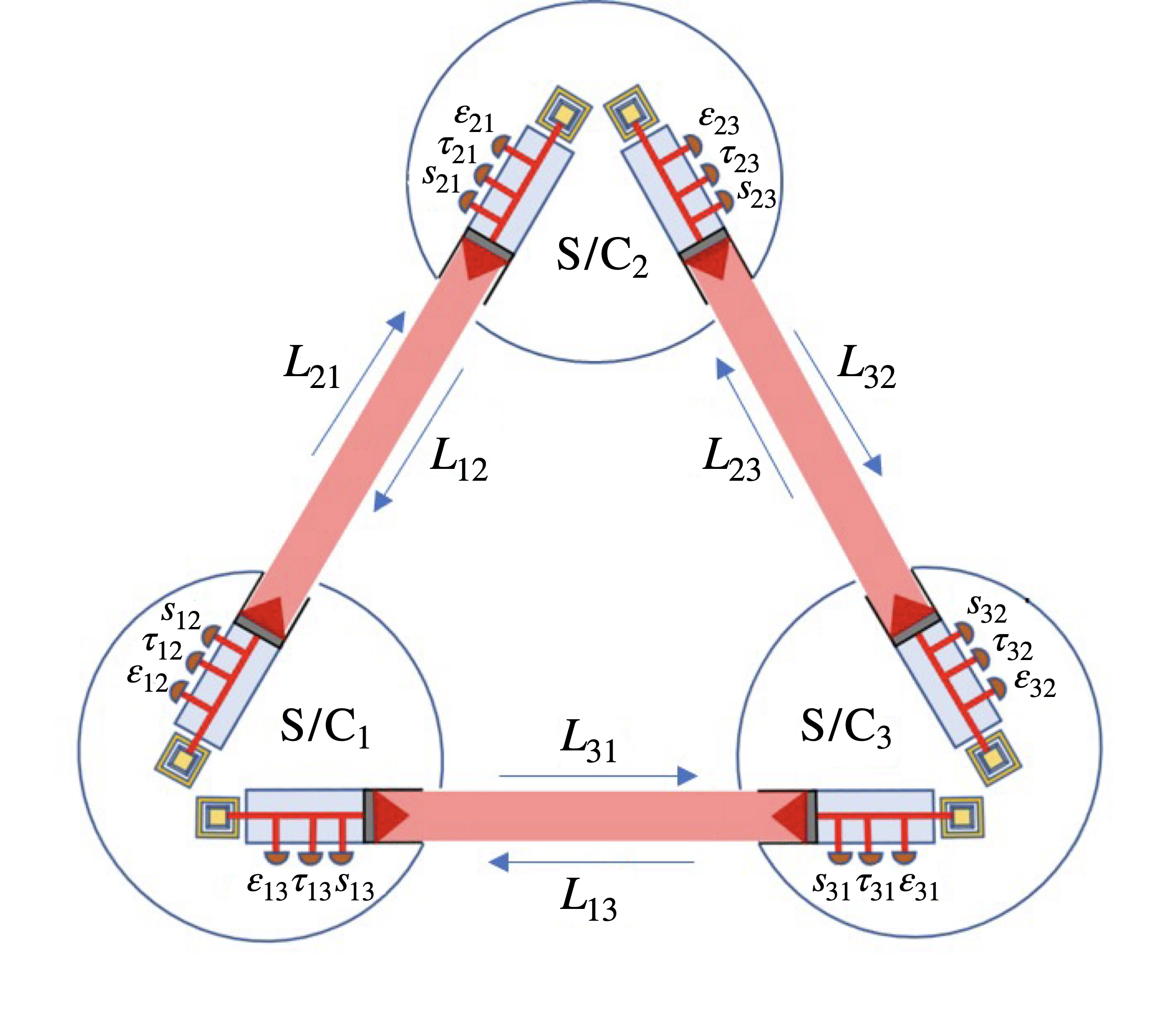}
    \caption{A schematic diagram of Taiji constellation. 
    Each SC is equipped with  two movable optical sub-assemblies (MOSAs), 
    which is a structure connecting  
    a telescope, a gravitational reference sensor (GRS) that {hosts a test-mass (TM)},  an  optical bench carrying the laser sources,  phasemeters, and other optical components needed for interferometric measurements, \emph{etc}.  
    Each MOSA is labeled by $ij$ ($ij \in \{12, 23, 31, 21, 32, 13\}$), with $i$ being the index of SC carrying this MOSA, and $j$ the index of distant SC that transmits lasers with this MOSA.
    The indexing of on-board interferometric measurements $\{ s_{ij}, \tau_{ij}, \varepsilon_{ij} \}$ are consistent with the MOSAs. 
    For laser link $L_{ij}$, {$i$ and $j$ denote} the SC that receives and emits laser, respectively. 
    }
    \label{fig:taiji_constellation}
\end{figure}

\section{Basic Mission Designs \& Scientific Data Flow of Taiji}\label{sec:2}

\subsection{Baseline mission design for Taiji}
As  schematically shown in FIG.~\ref{fig:taiji_constellation},  Taiji  consists of three SCs configured in an approximate equilateral triangle constellation, with nominal armlengths of $\sim 3\times 10^9$ m~\cite{taiji_0,taiji_1,taiji_2}. 
The constellation operates in a heliocentric orbit and leads the Earth by $\sim 20^\circ$. 

\begin{figure*}
    \centering
    \includegraphics[width=0.95\linewidth]{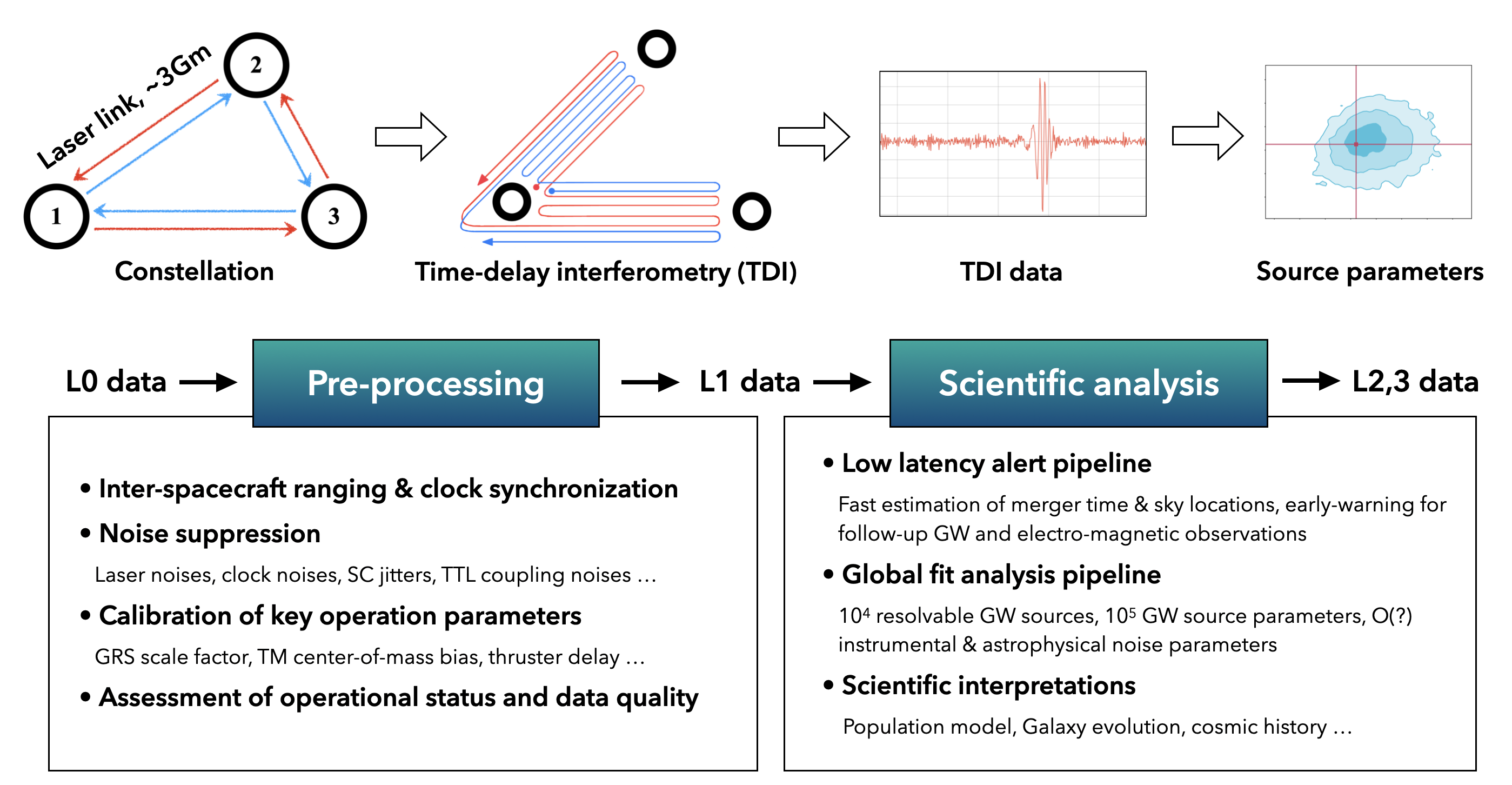}
    \caption{The scientific data flow of Taiji.}
    \label{fig:taiji_data_flow}
\end{figure*}

The basic concepts and requirements for Taiji's systems and payloads are outlined in Ref.~\cite{taiji_2}. 
In principle, the generation of mock  data should stay faithful to these designs. 
The core systems relative to TDC II include: 

    \textbf{$\bullet$ Laser interferometry system}: consisting of   $1064 \ {\rm nm}$-wavelength lasers with $30 \ {\rm Hz} /\sqrt{\rm Hz}$ frequency stability, $0.4 \ {\rm m}$-diameter telescopes with angular jitters of outgoing beams controlled at $1 \ {\rm nm} / \sqrt{\rm Hz}$ level, and Michelson-type interferometers achieving ${\rm pm} /\sqrt{\rm Hz}$ precision via heterodyne detection~\cite{taiji_interferometry_system}.
    In order to isolate the responses to GWs  from the jitters of SCs, Taiji adopts the ``split interferometry'' design, which separates the interferometric measurements into inter-spacecraft interferometers (ISIs, or science interferometers, dubbed $s_{ij}$, see FIG.~\ref{fig:taiji_constellation} for the indexing convention of TDC II), reference interferometers (RFIs, $\tau_{ij}$), and test-mass interferometers (TMIs, $\varepsilon_{ij}$). 
    The TM-to-TM measurements $\eta_{ij}$ used for GW detection are synthesized  from them  following two steps. Firstly,   
    \begin{equation}
    \xi_{ij} = s_{ij} + \frac{\tau_{ij} - \varepsilon_{ij}}{2} + {\textbf{D}}_{ij}\frac{\tau_{ji}-\varepsilon_{ji}}{2},
    \end{equation}
    where $ij \in \{12, 23, 31, 21, 32, 13\}$. 
    The delay operator is defined as  $\textbf{D}_{ij}f(t) \equiv f[t - d_{ij}(t)]$, $f$ being an arbitrary function of time, and $d_{ij}(t) \equiv L_{ij}(t) / c$ is the light travel time from SC$_j$ to SC$_i$. 
    Secondly, 
    \begin{equation}\label{eq:eta}
        \eta_{ij} = \xi_{ij} + {\textbf{D}}_{ij}\frac{\tau_{ji}-\tau_{jk}}{2}, \quad \eta_{ik} = \xi_{ik} + \frac{\tau_{ij}-\tau_{ik}}{2},
    \end{equation}
    with $ijk \in \{123, 231, 312\}$. 
    The sampling of the interferometric signals is triggered according to the onboard clocks, and the signals are  ultimately {measured by} the phasemeters. 
    Besides, the laser links also undertake functions such as clock noise transfering, inter-spacecraft ranging, and inter-spacecraft communications, \emph{etc}~\cite{ranging_communication,ranging}. 
    
    \textbf{$\bullet$ Gravitational reference sensor (GRS) and Drag-free \& altitude control system (DFACS)}: 
    Taiji employs  46-mm gold-platinum alloy TMs as  free-falling inertial references, which are shielded in the GRSs, in order to isolate the TMs  from  non-gravitational disturbances from space environment (\emph{e.g.} solar radiation pressure). 
    GRSs measure the  displacements of SCs relative to  TMs 
    with capacitive sensing resolution of $1.8 \ {\rm nm} / \sqrt{\rm Hz}$ along the sensitive axis and $3 \ {\rm nm}  / \sqrt{\rm Hz}$ along non-sensitive axes, 
    feeding the readouts to 
    the electrostatic servo control system of the GRS and the DFACS. 
    The GRS controller will force the TMs to maintain its nominal positions along the non-sensitive axes, while the DFACS 
    will commands $\mu$N thrusters 
    (force resolution 0.1 $\mu$N, noise $0.1 \ \mu {\rm N} / \sqrt{\rm Hz}$, and response time $< 0.33 \ {\rm s}$) 
    to ensure that SC follows the trajectories of TMs 
    along the sensitive axes.
    The residual displacement noise is required not to exceed ${\rm nm} / \sqrt{\rm Hz}$ level to reduce coupling stray forces. 
    Meanwhile, the pointing ahead angle mechanism (PAAM) actively compensates for deviations from line of sight caused by orbital dynamics, and stabilizes laser alignment with angular jitter at the  ${\rm nrad}  / \sqrt{\rm Hz}$ order, maintaining robust laser link continuity. 
    Under these controls, {Taiji} also requires that the  breathing angles to be within the scale of $\pm 0.5^\circ$ and the Doppler shifts between SCs within $\pm 5 \ {\rm MHz}$, during the whole mission lifetime.
    In principle, the full-scale simulation for  GRS and DFACS entails up to 60 degrees of freedom, spanning the magnitudes from pm to astronomical unit~\cite{dfacs_simulation}.

In the raw measurements  $\{s_{ij}, \tau_{ij}, \varepsilon_{ij}\}$, laser frequency noise, clock noise, and  SC jitter dominate over the target GW signals by orders of magnitudes. 
These ``primary'' noises will be suppressed via split-interferometry design and  the TDI data processing technique, 
thus the residual ``secondary noises'' ultimately determine the  sensitivity of Taiji. 
As a preliminary categorization, 
Taiji’s noise budget mainly  comprises two  components, one is the ``position noise'', which {includes} the contributions of multiple optical metrology system (OMS) noises, including shot noises and various optical path fluctuations, and the other is the test-mass acceleration (ACC) noise due to residual stray forces acted on the TMs. 
To  enable  GW observations in the 0.1 mHz - 1 Hz band, the requirements on Taiji's OMS and ACC noises are $A_{\rm OMS} = 8 \ {\rm pm} / \sqrt{\rm Hz}$ and $A_{\rm ACC} = 3 \ {\rm fm} / {\rm s}^2 / \sqrt{\rm Hz}$, respectively, with  spectral profiles designed as 
\begin{align}
    P_{\rm OMS}(f) &= A_{\rm OMS}^2 \left[1 + \left(\frac{2 \ {\rm mHz}}{f}\right)^4\right], \label{eq:OMS_PSD}  \\
    P_{\rm ACC}(f) &= A_{\rm ACC}^2   \left[1 + \left(\frac{0.4 \ {\rm mHz}}{f}\right)^2\right] \left[1 + \left(\frac{f}{8 \ {\rm mHz}}\right)^4\right], \label{eq:ACC_PSD}
\end{align}
where $P_{\rm OMS}(f)$ and $P_{\rm ACC}$ denote the power spectral densities (PSDs) of OMS and ACC noises, respectively.

\subsection{Taiji scientific data flow}
After the data are downlinked to the Earth, 
the {on-ground data processing flow transforms raw measurements into scientific outputs through  a two-stage procedure.
The design of this workflow draws inspiration from the research on Taiji, LISA and TianQin, and incorporates the experiences gained from  Taiji-1.}
{For clarity, we first introduce the definitions of  data levels used in this work: }

{\textbf{$\bullet$ Level 0 ($L_0$) data}: Raw, unprocessed measurements downlinked from the SCs.}

{\textbf{$\bullet$ Level 1 ($L_1$) data}: Data processed through  TDI to suppress the primary noises, resulting in TDI variables used by further scientific analysis.}

{\textbf{$\bullet$ Level 2 ($L_2$) data}: Output of the  global fit pipeline, providing  posterior parameter distributions for  candidate GW sources.}

{\textbf{$\bullet$ Level 3 ($L_3$) data}: Final catalog of GW sources, astrophysical and cosmological parameters.}

The two-stage processing workflow is as follows  (see FIG.~\ref{fig:taiji_data_flow} for a brief summary):

    \textbf{$\bullet$ The pre-processing stage} ($L_0$ data $\rightarrow$ $L_1$ data, \emph{i.e.} raw data to TDI outputs) aims to provide data that are suitable for scientific analysis, with primary noises effectively mitigated, and synchronized to a global time frame (\emph{e.g.} the Barycentric Coordinate Time of Solar system, TCB for short).
    {This stage includes inter-spacecraft ranging~\cite{PhysRevD.109.022004,ranging_communication,ranging,DU2024107819}}, clock synchronization~\cite{clock_sync,Reinhardt:2024plf}, {and the suppression of laser frequency noise~\cite{wtn_tdi,Armstrong_time_1999,Tinto_time_2021,Hartwig:2022yqw,Wang:2020pkk}}, 
    {clock noise~\cite{tdi_clock_noise_markus,tdi_clock_noise_tinto,tdi_clock_noise_olaf,taiji_clocknoise_experiment}}, SC jitter, tilt-to-length (TTL) coupling~\cite{ttl_noise_subtraction,ttl_noise_subtraction_farfield} and other noises~\cite{Wu:2022qov,PhysRevD.111.024011} during or after the TDI processing. 
    In a broader sense, it should also incorporate the calibration of key operating parameters  of the science payloads and measurement links  (\emph{e.g.} the GRS scale factors, the bias of TM center-of-mass~\cite{GRS_calibration_taiji1_haoyue,GRS_calibration_taiji1_xiaotong}, the thruster delay, \emph{etc}), as well as the evaluation of their performances and data quality.
    As a critical step within pre-processing, TDI mitigates the laser frequency noise, which is predicted to be 6-8 orders of magnitude stronger than GW signals, by delaying and combining the raw interferometric measurements to synthesize virtual equal-armlength interferometry. 
    According to the  ways of combination, TDI can be categorized into different configurations (\emph{i.e.} TDI channels),  such as the Michelson type, Sagnac type, Monitor type, Relay type, \emph{etc.}~\cite{Markus_PhD_thesis,Olaf_PhD_thesis};
    or different generations, such as the 1st generation designed for static constellations, 
    and the 2nd generation more suitable for the  flexing  armlengths of LISA and Taiji. 
    Regardless, all the TDI combinations can be written in a unified form:
    \begin{equation}
    {\rm TDI} \ = \sum_{ij} \textbf{P}_{ij} \eta_{ij},
    \end{equation}
    with $\textbf{P}_{ij}$ synthesized from delay operators. 
    For example, the rule of Michelson-$X_2$ channel  used in TDC II reads 
    \begin{align}
        \textbf{P}_{12} &= 1 - \textbf{D}_{131} - \textbf{D}_{13121} + \textbf{D}_{1213131}, \nonumber \\ 
        \textbf{P}_{23} &= 0, \nonumber \\ 
        \textbf{P}_{31} &= -\textbf{D}_{13} + \textbf{D}_{1213} + \textbf{D}_{121313} - \textbf{D}_{13121213}, \nonumber \\ 
        \textbf{P}_{21} &= \textbf{D}_{12} - \textbf{D}_{1312} - \textbf{D}_{131212} + \textbf{D}_{12131312}, \nonumber \\ 
        \textbf{P}_{32} &= 0, \nonumber \\ 
        \textbf{P}_{13} &= -1 + \textbf{D}_{121} + \textbf{D}_{12131} - \textbf{D}_{1312121},  
    \end{align}
    where $\textbf{D}_{i_1i_2i_3 ...} f(t) \equiv \textbf{D}_{i_1i_2}\textbf{D}_{i_2i_3}...f(t)$.\\

    \textbf{$\bullet$ The scientific analysis stage} ($L_1$ data $\rightarrow$ $L_{2,3}$ data, \emph{i.e.} TDI outputs to source parameters and scientific interpretations) starts from the TDI data, and aims to accomplish the detection of GW signals, the estimation of source parameters, and further interpretation of GW physics. 
    At least {two pipelines play} critical roles in this stage. 
    One is the ``low latency'' alert pipeline~\cite{low_latency_cornish,near_real_time_tianqin,AI_mbhb1,AI_mbhb2,AI_merger_prediction},  a real-time processing system designed to rapidly detect and locate transient GW events, enabling timely electromagnetic follow-up observations as well as potential 
    multi-mission (\emph{e.g.} LISA-Taiji~\cite{lisa_taiji_network1,lisa_taiji_network2}, LISA-Taiji-{TianQin}~\cite{lisa_taiji_tianqin_network1,lisa_taiji_tianqin_network2} networks) and 
    multi-band GW observations. 
    The other is the ``global fit'' analysis pipeline~\cite{Littenberg_prototype_2023,Strub_global_2024,Katz_efficient_2024}, which performs computationally intensive Bayesian analysis to disentangle overlapping GW signals, precisely estimating the parameters of  thousands of sources, and  generate high-confidence source catalogs~\cite{2025arXiv250214818J}.

\section{Challenges in Taiji Data Analysis}\label{sec:3}

{
{This section presents  systematic review of the anticipated challenges in future data analysis. }
Several demonstrations based on our simulation are provided in FIG.~\ref{fig:higher_mode_mbhb} - FIG.~\ref{fig:T_channel_sensivitity}. 
Given  the similarities between Taiji and LISA in terms of mission concept, target sources and sensitivity, 
the issues discussed here are relevant to both missions, 
and we expect this work to support the development of algorithms and tools with broad applicability across similar detectors.
}

\subsection{The challenges of signal modeling and computational demand}\label{subsec:3.1}

Matched filtering is one of the most classical and well established algorithms in GW data analysis~\cite{matched_filter_Davis1989ARO,matched_filter_Jaranowski_Krolak_2009}, and 
its implementation necessitates the accurate and efficient modeling of signals (\emph{i.e.} templates). 
{On the one hand, for a single GW event, Bayesian parameter estimation  typically involves more than $\mathcal{O}(10^6)$ template evaluations to sample high-dimensional posterior distributions~\cite{PhysRevLett.124.041102}. }
On the other hand, high accuracy is also crucial to ensure that the templates can precisely match the actual GW signals, thereby maximizing the detection efficiency and minimizing false positives and biased estimation.
These  demands are  also shared by
machine learning methods such as simulation-based inference~\cite{PhysRevLett.124.041102,SBI_NPE}, since {they rely on} the forward models of signals to generate large training sets.
The template for GW signal comprise two  aspects:  waveform (polarizations) and detector response (single-link response and TDI combination). 
Achieving both speed and accuracy  can be a challenging task, especially in the context of space-based GW detection.

\begin{figure}
    \centering
    \includegraphics[width=0.95\linewidth]{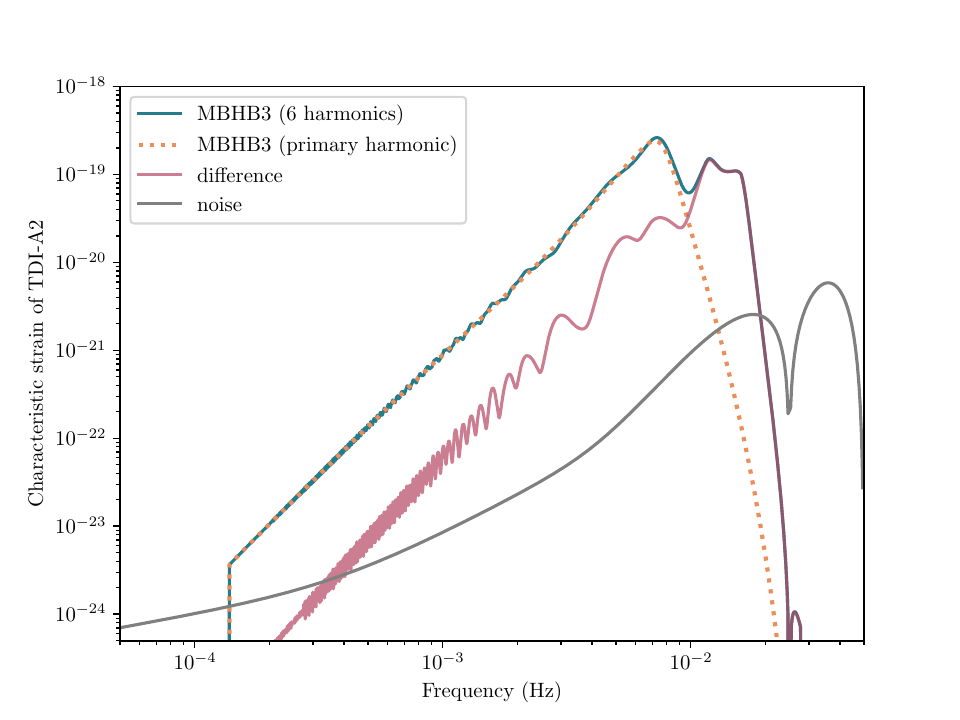}
    \caption{
    Detectability of the higher harmonics, showcased using the parameters of the 3rd MBHB from dataset 1.1. 
    The green solid curve represents the waveform (in terms of TDI-$X_2$ response) containing only the dominant (2,2) mode, while the orange dotted curve incorporates 6 harmonic modes: (2,1), (3,3), (3,2), (4,4), and (4,3). 
    Systematic deviations caused by neglecting higher modes are quantified by the residual (yellow line), compared against the instrument noise floor (gray curve) defining the detection threshold.
    This mismatch necessitate the inclusion of higher modes for unbiased parameter estimation in the high-SNR regime.}
    \label{fig:higher_mode_mbhb}
\end{figure}

A variety of quantitative criteria have been raised to assess and limit the accuracy of templates~\cite{PhysRevD.95.104004,PhysRevD.109.104043}, such as the  mismatch threshold~\cite{Lindblom:2008cm,PhysRevResearch.2.023151}, the parameter estimation error based on Fisher formalism~\cite{PhysRevD.76.104018}, \emph{etc}. 
A key observation is that higher-SNR signals generally  demand more precise  templates  to avoid biased  estimation, since for high SNRs, systematic  errors can easily dominate statistical uncertainties. 
Among the target sources of Taiji and LISA, MBHBs can achieve SNRs of $\mathcal{O}(10^3)$ and SNRs of $\mathcal{O}(10^2)$ are also foreseen for bright GBs. 
The overlap of signals  further exacerbates these requirements. 
For instance, the matching residual of one signal can potentially disrupt the accurate resolution of others. 
Meanwhile, this also indicates that subdominant features in the waveforms,  such as the eccentricity~\cite{10.1093/mnras/stad3477}, high harmonics beyond the dominant $(2,2)$ mode~\cite{PhysRevD.108.044053,Gong:2023ecg} (see FIG.~\ref{fig:higher_mode_mbhb} for an illustrative example), and precession~\cite{Pratten:2023krc} of MBHB, as well as the eccentricity of GB~\cite{10.1093/mnras/stae1288}, would become non-negligible.
These complexities pose both theoretical and computational challenges for data analysis, on the other hand, they could be beneficial for testing general relativity and explore GW astrophysics~\cite{Liu:2024jkj,PhysRevD.106.103017}.

\begin{figure*}[t]
    \centering
    \includegraphics[width=\linewidth]{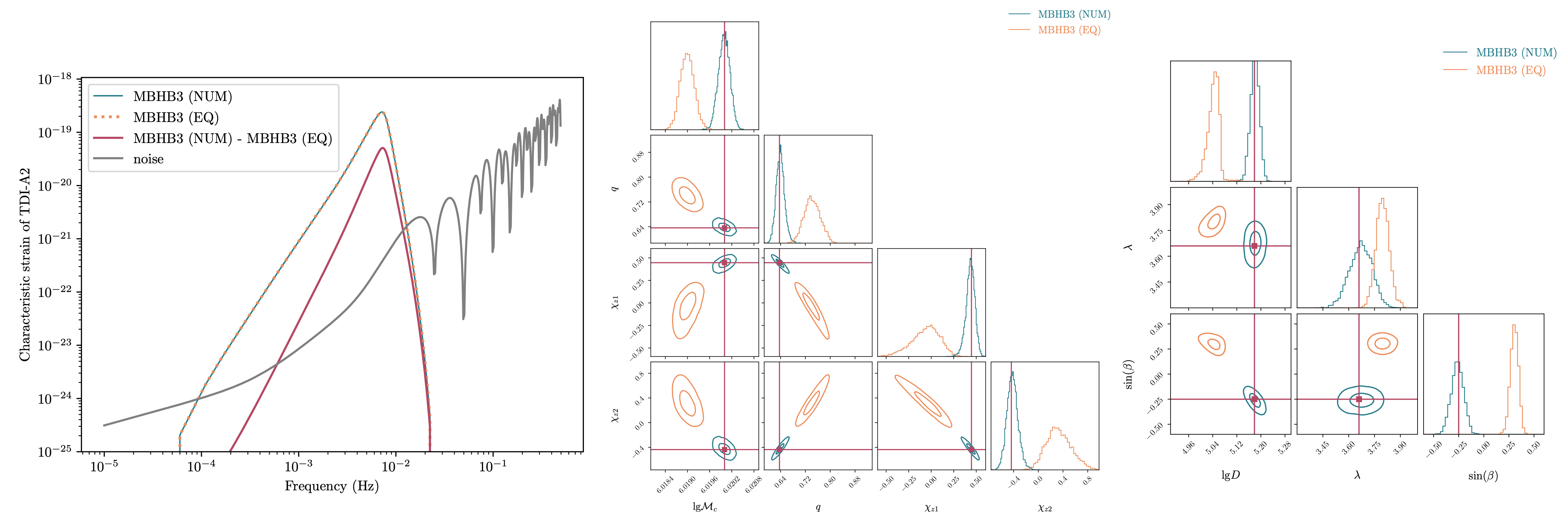}
    \caption{
    Systematic errors induced by inaccurate orbit model. 
    Using the same source parameters as FIG.~\ref{fig:higher_mode_mbhb}  as an example, the data of TDI-$X_2$ channel is simulated in the time domain, with GW response calculated based on a numerical orbit.
    Notice that we have excluded instrumental noises to isolate systematic biases from statistical uncertainties. 
    When estimating the MBHB's source parameters,  two response templates are employed:  (1) faithfully adopting the same numerical orbit, and (2) using an analytical equal-armlength approximation closest to the numerical orbit. 
    The left panel illustrates the discrepancy between these  two templates,  
    while  comparisons on the estimates for intrinsic  and extrinsic (3D localization) parameters are presented in the middle and right panels, respectively. 
    Evidently, for MBHBs with SNRs exceeding $10^3$, the incorporation of realistic orbital information  is critical to  avoid significant systematic biases.}
    \label{fig:mbhb_orbit_systematic_error}
\end{figure*}

For  space-based detections, the detector responses to GW signals diverge form those of LVK in at least two aspects. 
{Firstly, the rotation and translation of detector constellation induce time-dependent antenna pattern functions and  Doppler effects, respectively. 
While these  effects complicate response modeling,  they also help to alleviate the degeneracies among extrinsic parameters, thereby enhancing the accuracy of source localization~\cite{Marsat:2020rtl}.}
Secondly, the configurations of TDI combination, as  well as the armlength information (used as delays) also shape the response formalism. 
Current  studies, particularly those targeting LDC, predominantly utilize 1st-generation Michelson  TDI configurations under the equal-armlength approximation. 
However, these frameworks become insufficient when applied to  realistic Taiji / LISA orbits, where armlengths vary at rates up to several  ${\rm m}/{\rm s}$. 
Therefore it is imperative to employ 
2nd-generation TDI configurations and account for dynamic armlengths. 
 Ref.~\cite{Katz2022Assessing} demonstrated that, for high-SNR GBs ($\mathcal{O}(10^2)$), using the idealistic equal-armlength analytic orbit in response modeling would induce  significant systematic biases in parameter estimation.
Applying the same investigation to  MBHBs with SNR $\mathcal{O}(10^3)$,  the resulting systematic deviations for   both intrinsic and extrinsic parameters exceed the $3\sigma$ thresholds (see FIG.~\ref{fig:mbhb_orbit_systematic_error}).



Except for the stringent accuracy requirements, 
long-duration GW signals pose significant computational challenges due to the large data size
(\emph{e.g.} $\mathcal{O}(10^6)$  data points per year for a sampling rate of 0.1 Hz). 
Although semi-monochromatic signals (\emph{e.g.} GBs)  can  be efficiently calculated in the frequency-domain~\cite{fastgb}, chirping signals with time-frequency evolution, such as sBHB inspirals in the mHz-band which  undergo significant frequency evolution over years, {render traditional  methods inadequate~\cite{sBHB_wavelet,sbhb_tianqin,sbhb_tianqin_ai,Fu:2024cpu}.}
The EMRI waveforms, on the other hand, besides sharing the computational bottlenecks with sBHBs in template volume, face unique challenges due to their extreme complexity.
To extract the wealth of GW science from observation, a sub-radian accuracy in phase (more specificly $\Delta \Phi < 1 / {\rm SNR}$)~\cite{emri_sub_radian_requirement1,emri_sub_radian_requirement2} is required. 
While, the fast calculation of EMRI waveforms is hindered by their $\mathcal{O}(10^3)$ to $\mathcal{O}(10^5)$ harmonic modes. 
Current acceleration strategies, including semi-relativistic kludge models~\cite{kludge1,kludge2,kludge3,kludge4}, GPU parallelization, and deep-learning techniques~\cite{fastemri1,fastemri2}, remain limited by incomplete incorporation of  self-force corrections and post-adiabatic trajectories~\cite{emri_self_force}. 
Additionally, 
parameter estimation {is further complicated by}  multiple local maxima and  significant  ridges  in the likelihood~\cite{emri_local_maxima}.
Driven {by its prospects} to probe astrophysics~\cite{emri_astrophys}, cosmology~\cite{emri_cosmology} and gravity theory~\cite{emri_gravity_test}, ongoing  research will keep focusing  on developing fast, accurate  templates and  robust statistical inference frameworks~\cite{Yun:2023vwa,Ye:2023lok}.

Additionally, GWs reaching the detectors are further influenced by both the sources' local environments and the GWs' propagation paths. 
Omission of these elements in the waveform models might also introduce bias in the analysis.
For the former, examples include effects reviewed in Ref.~\cite{Chen2020}, including the relative motion of the source, the presence of a nearby massive object, and a gaseous background, {\emph{etc},  
while the latter involves} effects such as gravitational lensing~\cite{PhysRevLett.105.251101,Takahashi2003iWave,Lin2023Detecting,Cao2014Gravitational}.
For example, LISA is predicted to detect several multiply imaged GW signals during its mission lifetime~\cite{PhysRevLett.105.251101}, and the estimation is similar for Taiji. 
These phenomena also offer opportunities to probe the  structures of lens objects, test galactic models, and constrain cosmic evolution history. 

\subsection{The challenges due to noises and anomalies}\label{subsec:3.2}
\begin{figure*}[t]
    \centering
    \includegraphics[width=0.9\linewidth]{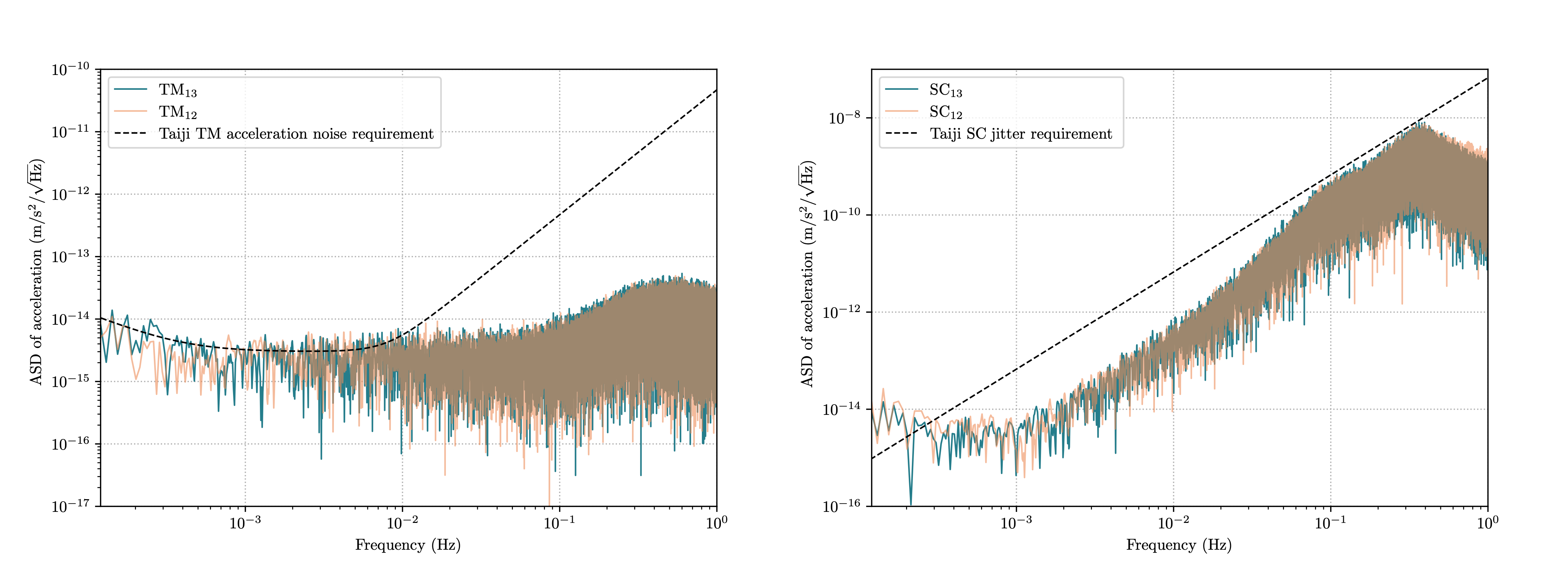}
    \caption{
    The TM ACC noises (left panel) and SC  jitter noises (right panel) derived  from the numerical simulation for the DFACS. 
    {Each noise component meets Taiji's design sensitivity} requirements (black dashed curves), yet none of them are fully aligned with  the  design curves.}
    \label{fig:DFACS_noise}
\end{figure*}

Accurate characterization of  noises is crucial not only for the  
 analysis of resolvable signals, but also for distinguishing  among SGWB, confusion foreground and instrumental noises~\cite{PhysRevD.89.022001, SGWBinner_Caprini_2019,SGWBinner_Flauger:2020qyi,Savalle2022Assessing}. 
However, unlike the LVK observatories which utilize auxiliary data channels from various sensors for real-time noise  monitoring, calibration and isolation~\cite{LIGO_noise_monitor,VIRGO_noise_monitor}, 
due to its space-based mission design, 
Taiji lacks equivalent  diagnostic methods. 
Furthermore, the distinct  environments of  ground-based experiments and in-orbit {operations suggest} that  the statistical properties of  noise might not be completely  determined prior to launch~\cite{unknown_noise,gaussian_process_noise}. 
FIG.~\ref{fig:DFACS_noise} displays the TM ACC noises (left panel) and SC  jitter noises (right panel) derived  from a numerical simulation for the  DFACS~\cite{dfacs_simulation}. 
As  shown, each noise component {meets Taiji's design} sensitivity requirements (black dashed curves), yet none of them are fully aligned with  the  design curves. 
These discrepancies, together with the signal-dominated regime, highlight the necessity of 
joint parameter estimation on signals and noises, using the limited  data transmitted to Earth~\cite{prototype_sgwb_global_fit,spectra_separation}.
The agnostic of noise spectral profile requires flexible modeling as well as trans-dimensional search in unknown parameter space~\cite{gaussian_process_noise,unknown_noise,bayesline} 

\begin{figure}
    \centering
    \includegraphics[width=0.95\linewidth]{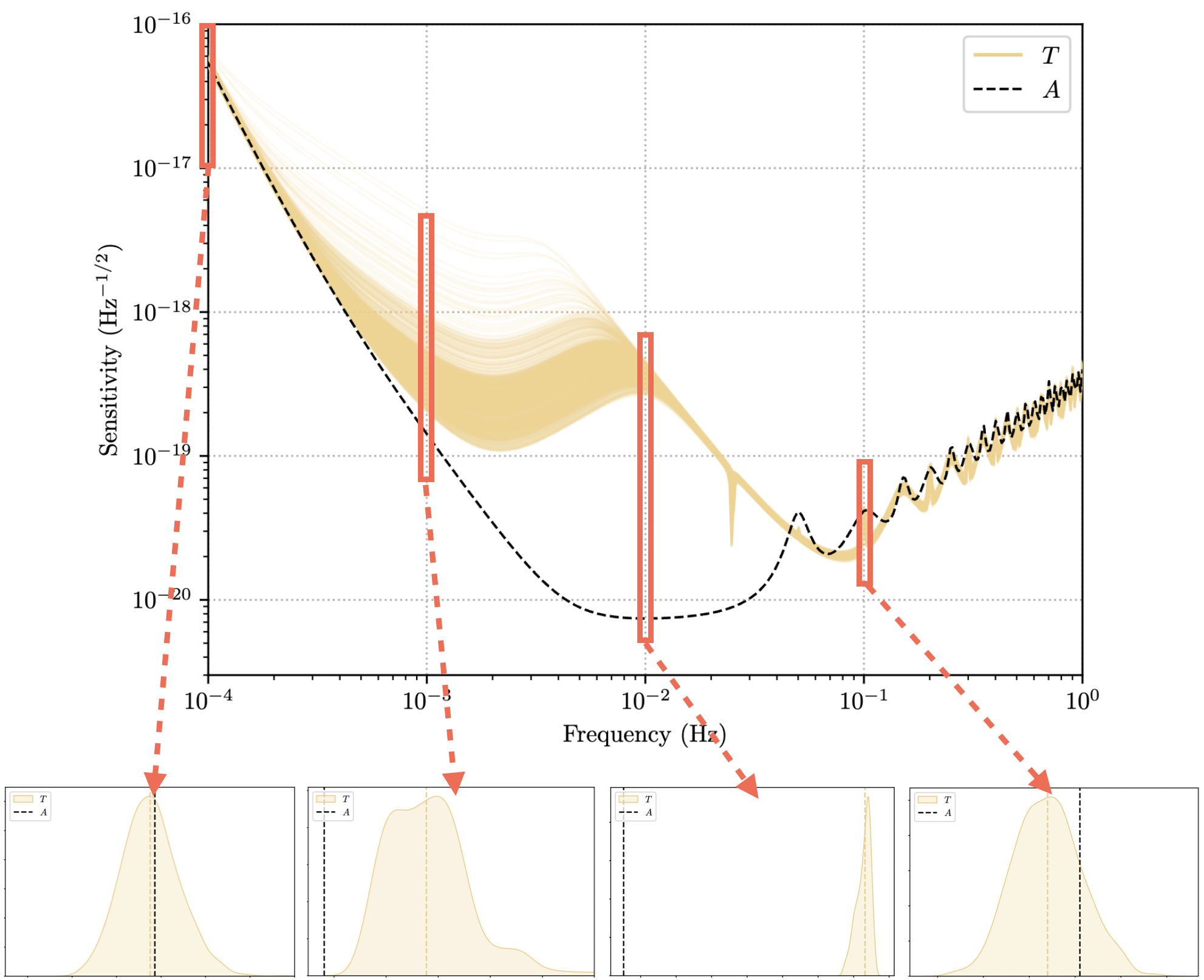}
    \caption{
    The sensitivity of  Michelson-$T_2$ (yellow curves) channel across a year, compared to the ``signal'' channel Michelson-$A_2$ (black dashed curve). 
    We have selected four representative frequencies, whose  sensitivity  distributions are displayed in the three subplots below. 
    }
    \label{fig:T_channel_sensivitity}
\end{figure}

Additionally,  {noise characteristics also inherently depend on} the TDI configurations. 
{The  widely adopted Michelson $\{A, E, T\}$ channels are  usually referred to as the ``optimal'' TDI combinations in the literature~\cite{TDI_optimal_channels}. 
Under the assumptions of equal armlength and identical noise spectra for each MOSA and each laser link, the optimal combinations exhibit  advantages such as noise orthogonality, which  enables direct summation of the statistics (SNRs, log likelihoods, \emph{etc}) across individual channels.
Besides, the $T$ channel, being insensitive to GW signals at low frequencies, can be particularly valuable as a ``null'' channel used for   instrumental noise characterization  and  SGWB detection.  
However, for realistic orbits with unequal and time-varying armlengths, the optimal channels will have non-negligible noise covariance~\cite{unknown_noise},  thus practical data analysis must account for the covariances among channels, increasing both modeling and computational complexities. 
The ``null'' property of $T$ channel will also be compromised  at both low and high frequencies~\cite{unequal_arm_tdi}.
FIG.~\ref{fig:T_channel_sensivitity} shows the sensitivity of  Michelson-$T_2$ channel across a year, with different yellow curves represent the sensitivities on different days, compared to the ``signal'' channel Michelson-$A_2$~\footnote{{In principle, PSD can effectively describe the statistical properties of noise only under the condition that the noise is Gaussian and stationary. Here the variation of the TDI noise profile is induced by armlength variation, which has a timescale of month (see FIG.~\ref{fig:LTTs_and_DPLs}). Therefore, for the observation of transient signals with durations on the order of days (such as MBHB merger), the noise can still be treated as locally stationary. This allows for the use of PSD and the calculation of sensitivity within the relevant time intervals of interest.}}. 
Consequently, the presumed optimality of $\{A, E, T\}$ 
demands further validation, 
and a systematic exploration on alternative TDI combinations remains imperative.}
Under different scenarios, various TDI schemes with specialized advantages have been developed, \emph{e.g.}, TDI channels suitable for noise characterization~\cite{Muratore2021Time, Muratore2022Effectiveness, Wang2022Characterizing, Hartwig2022Characterization,WANG2024107481}, robust against armlength variation~\cite{GangWang_time_2024, Wang:2024hgv},  
minimizing the loss due to data gap~\cite{Wang:2025mee},
 designed to reduce  specific noises~\cite{PhysRevD.111.024011, Wu:2022qov}, \emph{etc}.

Current studies on space-based GW data analysis usually assume Gaussian stationary noise and continuous  data streams.
While,  practical in-orbit operations would inevitably encounter non-ideal conditions such as glitches, data gaps, and non-stationary noises.
Firstly, glitch is a representative form of non-Gaussian and transient noises.  
Documented  in the data of  GRACE~\cite{Peterseim2014TWANGSH}, LISA Pathfinder~\cite{Armano2022Transient}, Taiji-1~\cite{Wu:2022qov}, 
these transients are also anticipated for  the future Taiji / LISA  missions.
Glitches could be harmful for scientific analysis by  inducing false alarms~\cite{Robson2019Detecting}, missed detections~\cite{Spadaro2023Glitch}, and parameter estimation biases~\cite{Castelli:2024sdb}. 
Secondly, the occurrences of data gaps, either  scheduled ones due to maintenances or maneuvers (\emph{e.g.} antenna repointing, frequency plan adjustments, PAAM rotation, \emph{etc}), 
or unscheduled ones due to in-orbit disruptions, are also expected~\cite{lisa_red_book,mind_the_gap}, {resulting in an effective  duty cycle of $\sim 75\%$~\cite{lisa_mission_duration}. }
Data gaps could degrade scientific outcomes through  breaking the coherence of long wave trains, insufficient SNR accumulation, increased noise floor caused by spectral leakage,  {or inducing  extra uncertainty in parameter estimation}~\cite{Carre2010The,Baghi2019Gravitational,Dey2021Effect,Castelli:2024sdb}.
Thirdly, non-stationary noises might emerge from various mechanisms. 
A guaranteed contributor is the cyclostationary astrophysical foreground caused by the  anisotropic distribution of GBs~\cite{PhysRevD.89.022001,Lin:2022huh}. 
Other mechanics include instrumental noise drifts due to the variations of payloads and SC platforms  during the multi-year mission lifetimes, as  evidenced in LISA Pathfinder~\cite{Armano2016SubFemtoG,Armano2018Calibrating,Armano2018Beyond}. 
This time-varying noise necessitates real-time characterization and calibration. 
Crucially, {non-stationarity} would  violate the fundamental  assumptions of conventional Whittle likelihood framework~\cite{whittle_likelihood},
significantly increasing computational demands for Bayesian inference~\cite{Cornish2020Time}.



\subsection{The ``global fit'' challenge}

The data of space-based GW detection is often metaphorically described as a cocktail, since it is a mixture of noises and numerous overlapping signals.
The sensitive band of Taiji and LISA contains tens of millions of GW sources, each with duration up to months or years. 
Therefore signal overlapping in both time domain and frequency domain is ubiquitous. 
For data $d= n + h^{\rm true}_{\rm total} = n+\sum_i h^{\rm true}_i$, assuming that the statistical property of noise $n$, as well as the types and numbers of signal templates $h_i^{\rm tem}$ are known (although not in practice), the full logarithmic likelihood can be written as 
\begin{align}
    {\rm ln}\mathcal{L} &= -\frac{1}{2} \left( d - h^{\rm tem}_{\rm total} | d - h^{\rm tem}_{\rm total} \right) \nonumber \\ 
    &= -\frac{1}{2} \left(\sum_i h^{\rm tem}_i \bigg| \sum_j h^{\rm tem}_j\right) + \left( \sum_i h^{\rm tem}_i \bigg| d \right) + {\rm const.} \nonumber \\ 
    &= -\frac{1}{2} \sum_i \left(h^{\rm tem}_i | h^{\rm tem}_i\right)  + \sum_i \left(h^{\rm true}_i \big| h^{\rm tem}_i \right)  + \sum_i \left( n | h^{\rm tem}_i\right) \nonumber \\
    & \quad - \sum_{i<j} \left(h^{\rm tem}_j | h^{\rm tem}_i \right) + 2\sum_{i<j} \left(h^{\rm true}_j \big| h^{\rm tem}_i\right)  + {\rm const.} \nonumber \\ 
    &\neq \sum_i {\rm ln}\mathcal{L}_i,
\end{align}
{where $\mathcal{L}_i$ is the likelihood function defined by comparing the $i$th individual signal to corresponding template ( \emph{i.e.} ${\rm ln}\mathcal{L}_i \equiv -1/2(h_i^{\rm tem}|h_i^{\rm tem}) + (h_i^{\rm true}|h_i^{\rm tem}) + {\rm const.}$).} 
Namely, due to the coupling between signals, estimating all the source parameters requires simultaneous modeling of the overlapping signals, rather than analyzing in an individual or sequential manner.
In other words, imperfect waveform subtraction risks residual contamination and biased subsequent analyses. 
Meanwhile, the detectability of a signal is determined not only by its intrinsic SNR, but also by the coupling with other signals.
Global fitting generally involves searching $\mathcal{O}(10^5)$ parameters across the parameter spaces of signals and  noises, hence creating computational bottlenecks, given that full Bayesian analysis of a single signal using nested sampling~\cite{Jonathan_nested_sampling} or Markov chain Monte Carlo (MCMC)~\cite{Cornish_mcmc} is already a time consuming task. 
Besides, overlapping signals with similar  morphologies (especially GBs) further  requires extra algorithms and computational powers for trans-dimensional sampling or  model selection~\cite{fastgb}. 
At last, except for signals, instrument noises, SGWBs and confusion foregrounds  
all manifest as stochastic components of data, further complicating noise characterization and source separation.

In order to develop algorithms to resolve this mixture, efforts (\emph{e.g.} Refs.~\cite{Littenberg_prototype_2023,Strub_global_2024,Katz_efficient_2024,gb_xhzhang,gb_pgao,gee_moo,prototype_sgwb_global_fit}) have been made on earlier mock datasets (MLDC / LDC), demonstrating that it is possible to jointly estimate  the parameters of signals and noises. 
These prototype pipelines typically demands week(s) of computation, where  the employed or developed  techniques include   F-statistics, parallel tempering MCMC (PTMCMC)~\cite{ptmcmc}, reversible jump (or trans-dimensional) MCMC~\cite{rjmcmc}, particle swarm optimization~\cite{pso}, differential evolution optimization~\cite{differential_evolution}, as well as fast and parallelizable (on CPU / GPU) waveform calculators~\cite{bbhx1,bbhx2,gbgpu1,gbgpu2,Katz2022Assessing}, \emph{etc}.
It should be noticed that, as discussed in Section~\ref{sec:1}, these datasets were built on  idealistic assumptions, including equal-armlength analytic orbits, simplified waveform models, mildly overlapping signals, Gaussian stationary noises with known spectral shapes. 
Therefore, global fit remains the core challenge for space-based GW data analysis, and current methodologies must be rigorously re-examined and extended in more realistic scenarios.




\subsection{The coupling issues between pre-processing and scientific analysis}
The coupling between pre-processing (noise suppression, operating parameter calibration, \emph{etc}) and scientific analysis (signal detection, source parameter estimation \emph{etc}) arises due to their shared dependence on the interferometer readouts.
The dominating  bright GW signals may 
impede noise and instrument calibration, while imperfect  noise suppression or inaccurate noise characterization in turn degrades the reliability of GW source inference based on the residual data.
Under this topic, identified  challenges in  existing researches include TDI ranging (a data driven approach for inter-spacecraft ranging), the estimation and subtraction of TTL noises~\cite{Littenberg_prototype_2023,ttl_gw_signal}, \emph{etc}.
These challenges underscore the  necessity of  developing more comprehensive end-to-end pipelines, {and demand} the incorporation of all  these signals, noises and instrumental effects   into  simulation.

\section{The TDC II Datasets}\label{sec:4}

\subsection{Data simulation pipeline}\label{subsec:4.1}

\begin{figure*}
    \centering
    \includegraphics[width=0.9\linewidth]{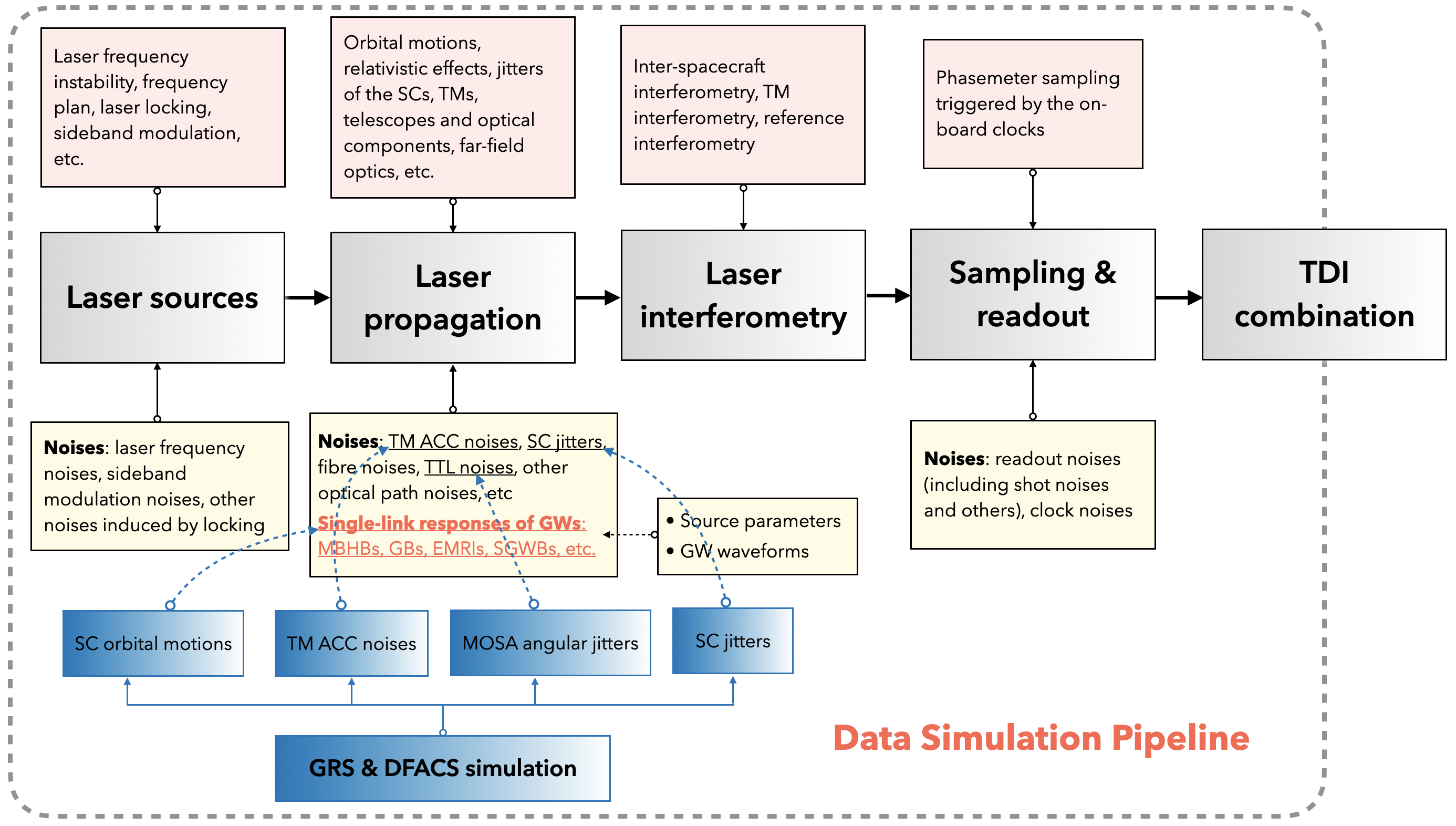}
    \caption{A concise view of the  data simulation pipeline.
    The grey branch represents the simulation framework of  laser interferometry measurements, with the upper red blocks denoting the physical processes and factors contributing  to each stage, 
    while the lower yellow blocks denoting the noises and GW signals newly introduced at each  stage.
    The simulations of GRS and DFACS as well as their outputs are schematically shown with the blue branch (to be detailed in separate publications). 
    }
    \label{fig:Taiji_mock_data_pipeline}
\end{figure*}

\begin{figure*}
    \centering
    \includegraphics[width=0.75\linewidth]{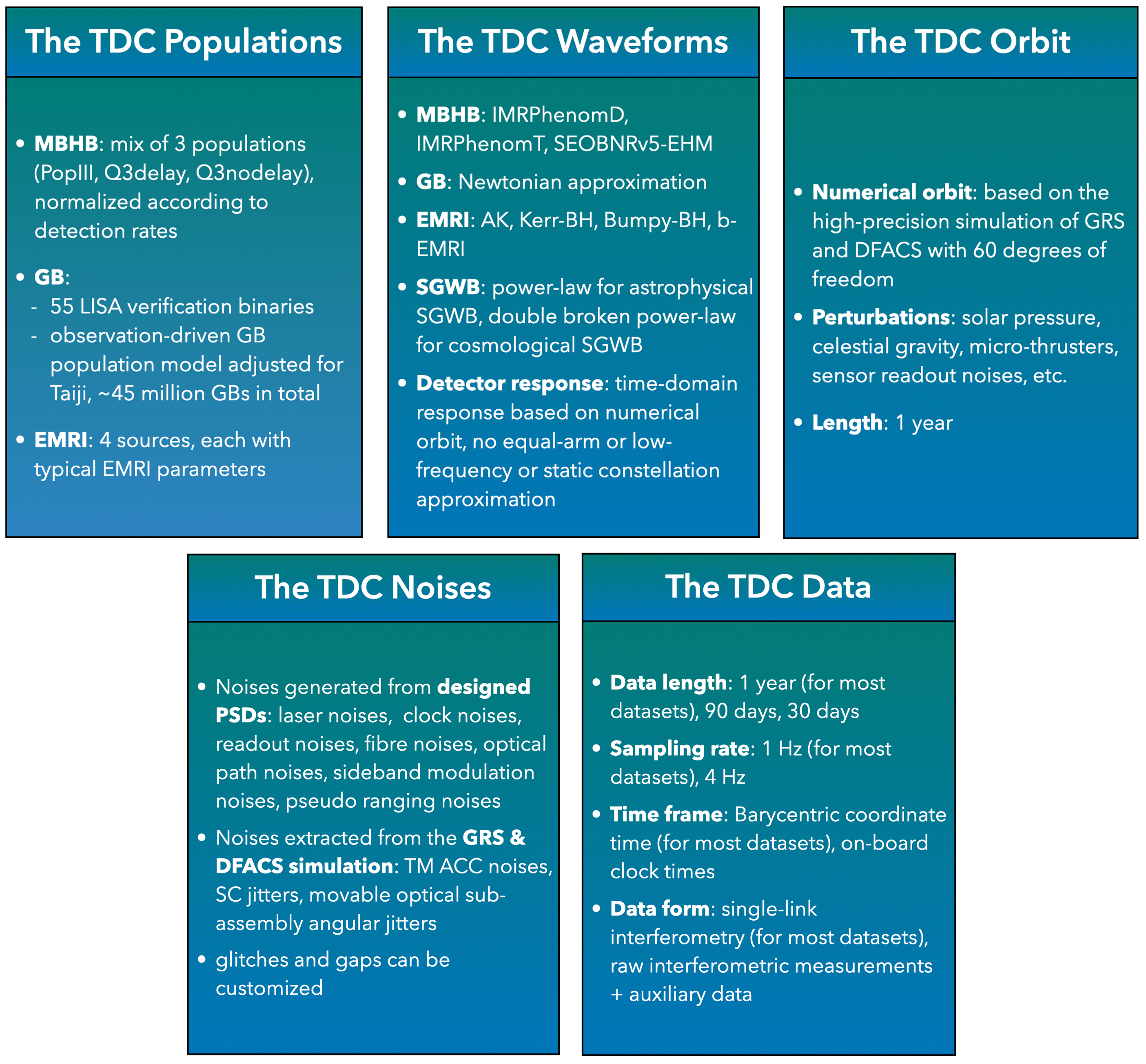}
    \caption{The models behind TDC II.}
    \label{fig:tdc_models}
\end{figure*}

    \begin{figure}
        \centering
        \includegraphics[width=0.95\linewidth]{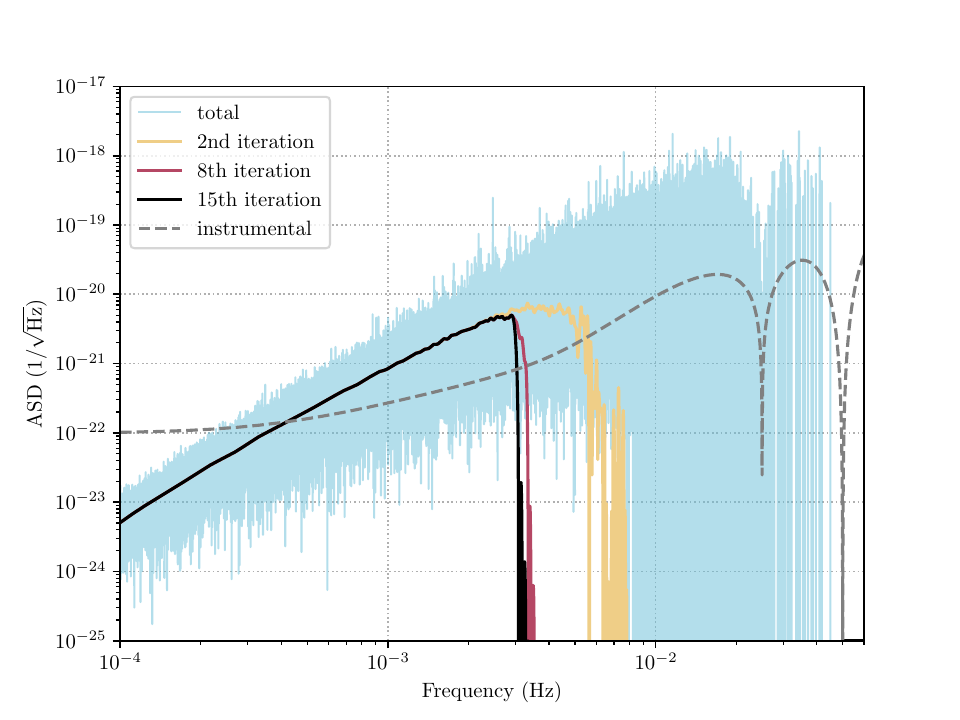}
        \caption{
        Preliminary estimation for the shape of confusion foreground. 
        The calculation is based on SNR threshold of 7 and 1-year observation time (in consistence with the datasets).
        We run the subtraction algorithm~\cite{gb_iterative_subtraction,taiji_confusion_noise} for 15 iterations to ensure convergence. 
        }
        \label{fig:GB_iterative_subtraction}
    \end{figure}

    \begin{figure}
        \centering
        \includegraphics[width=0.85\linewidth]{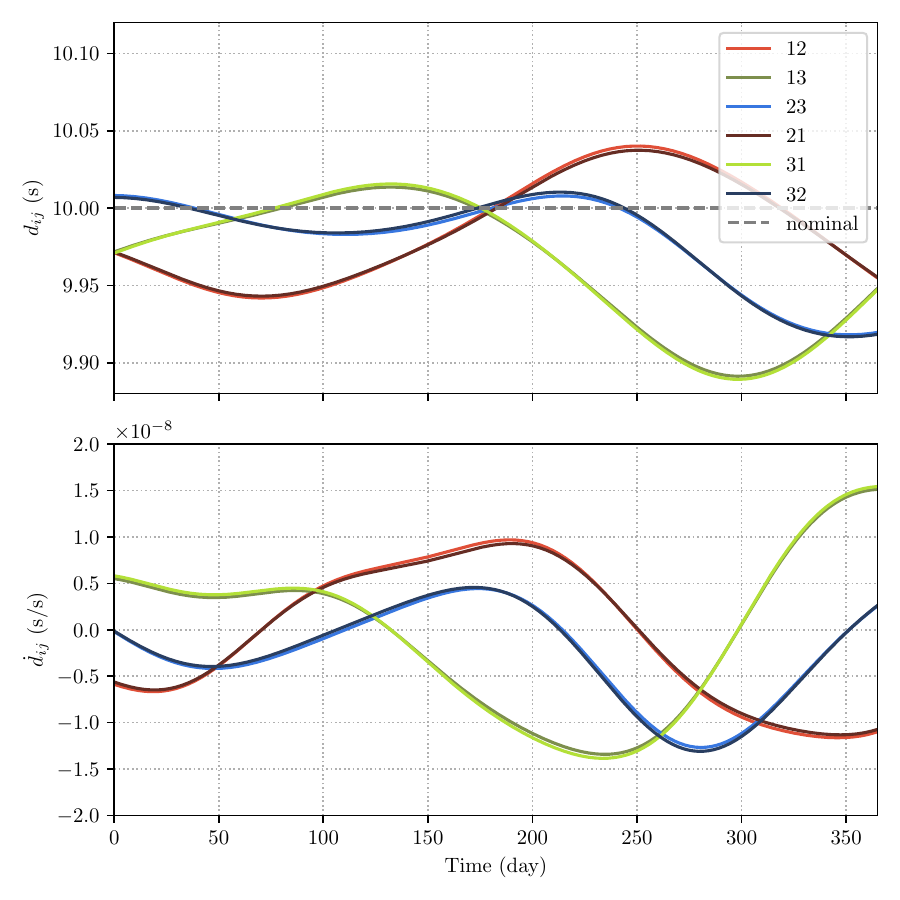}
        \caption{The light travel times (upper panel) and their derivatives (lower panel) along 6 arms $ij \in \{12, 23, 31, 21, 32, 13\}$  calculated according to the simulated Taiji orbit.
        The grey dashed line represents the nominal armlength (10 s, \emph{i.e.} $3\times 10^9$ m) of Taiji. 
        During the mission time of one year, the orbit has a maximum armlength difference of $\mathcal{O}(1 \%)$ and maximum armlength variation rate of less than 5 m/s.  }
        \label{fig:LTTs_and_DPLs}
    \end{figure}

{As  introduced in Section~\ref{sec:2}}, Taiji's data flow incorporates sophisticated in-orbit measurements and on-ground processing steps. 
In order to simulate  the performances of instruments and the characteristics of data within a reasonable timescale, we currently focus on the key information  {propagating through} the system, rather than implementing a full physical simulation. 
For Taiji's scientific data which mainly originate from laser interferometric measurements, we outline the framework of modeling and simulation as: 
laser sources $\rightarrow$ laser propagation $\rightarrow$ laser interferometry $\rightarrow$ signal sampling triggered by the on-board clocks $\rightarrow$ phasemeter readout $\rightarrow$ TDI processing and scientific analysis.
On the other hand, the simulations for GRS and DFACS constitute critical inputs to the aforementioned framework (to be detailed in separate publications). 
Specifically, provided by these simulations are the orbital motions of SCs, TM ACC noises, SC jitters, MOSA angular jitters, \emph{etc}.

In FIG.~\ref{fig:Taiji_mock_data_pipeline},  
the grey branch represents the simulation framework of  laser interferometry measurements, with  
the upper red blocks denoting the physical processes and factors contributing  to each stage, 
while the lower yellow blocks denoting the noises and GW signals newly introduced at each  stage.
The simulations of GRS and DFACS as well as their outputs are schematically shown with the blue branch. 
Notice that we have neglected  the downsampling step (from the  sampling rate of phasemeter, typically $\mathcal{O}(10)$ - $\mathcal{O}(10^2)$ MHz~\cite{taiji_phasemeter,taiji_multichannel_phasemeter},  to the sampling rate used for telemetry and data processing, $\mathcal{O}(1)$ - $\mathcal{O}(10)$ Hz) 
and directly simulate at the sampling rate of  final output data.
This simplification aims to avoid the  computational and storage cost caused by the massive volume of raw data.  

By design, the phasemeter of Taiji can output both phase and instantaneous frequency of interferometric signals~\cite{lisa_phasemeter_core,taiji_phasemeter,taiji_multichannel_phasemeter}.
In principle, data expressed in these two units are equivalent and can be  easily transformed to each other via differentiation or integration. 
{Using frequency unit is generally more convenient in terms of storage, telemetry and simulation,} since it remains at the same order of magnitude through the whole mission lifetime, and can be represented as double-precision floating-point numbers~\cite{lisa_instrument}. 
Therefore we simulate laser interferometry with frequency unit, and the outputs are further converted to fractional frequency differences by dividing the central frequency of laser (281.6 THz), to align with the convention of most literatures.

Within this framework, we list the models utilized in the creation  of TDC II as follows (also see FIG.~\ref{fig:tdc_models} for a more brief summary). 
Readers may refer to  \textbf{TDC II Manual}~\cite{TDCII_manual} for the mathematical formalism  of these models, as well as  their code implementations in \texttt{Triangle}.

    \textbf{$\bullet$ Source populations}: For MBHBs, we randomly draw the source parameters from the mix of 3 population models (PopIII, Q3delay, Q3nodelay)~\cite{mbhb_population_basemodel}, each   normalized according to the predicted event rates~\cite{mbhb_detection_rates}.
    The GB population includes 55 LISA verification Galactic binaries (VGBs)~\cite{VGB_gitlab,VGB_paper} (or more precisely, 55 ``detectable'' Galactic binaries, as  defined in Ref.~\cite{VGB_paper}), and $\sim 4.5 \times 10^7$ GBs generated from an 
    observation-driven population model~\cite{observation_driven_dwd_population} adjusted for Taiji (we expanded the  lower limit of GW frequency from 0.1 mHz to 0.05 mHz due to Taiji's better sensitivity at low frequencies). 
    To offer a preliminary estimate on the number of detectable GBs within this population and the spectral shape of confusion foreground, we employ the iterative subtraction method presented in  Refs.~\cite{gb_iterative_subtraction,taiji_confusion_noise}.
    Setting a widely adopted SNR threshold of 7 and observation time of 1 year, the theoretical upper limit for detectable GBs is $\sim 1.9 \times 10^4$, and the ASD of foreground is shown with black curve in FIG.~\ref{fig:GB_iterative_subtraction}.
    Besides, for the  4 injected EMRI signals, typical values of the source parameters are adopted.

    \textbf{$\bullet$ Waveforms}: For MBHBs, three different  approximants are adopted for different challenge topics: IMRPhenomD~\cite{imrphenomd,pycbc}, 
    IMRPhenomT~\cite{phenomt1,phenomt2,pycbc},
    and SEOBNRv5EHM~\cite{seobnrv5,pyseobnr}. 
    As for GBs, we adopt the  waveform under Newtonian approximation and expand the phase evolution up to the second derivative of GW frequency. 
    The injected EMRI waveforms include the augmented kludge (AK) model~\cite{AK_emri}, the EMRI waveforms of Kerr black hole (BH) and Bumpy BH~\cite{kerr_bumpy_emri,kerr_bumpy_emri1}, as well as the b-EMRI waveform, which is the EMRI system produced by tidal capture of binary black holes~\cite{bemri1,bemri2}. 
    Besides, we model the astrophysical origin SGWB with a power-law spectrum, and cosmological SGWB (first-order phase transition) with a  double broken power-law spectrum~\cite{prototype_sgwb_global_fit}.
    The detector responses are calculated in  time-domain response based on the rigorous formalism known so far~\cite{Katz2022Assessing,LDC_Radler_manual_v2,TDCI} (\emph{i.e.} no equal-armlength or low-frequency or static constellation approximations).
    
    \textbf{$\bullet$ Orbit}: The dynamics of SCs serve as critical inputs for both GW response calculation and noise simulation. 
    The numerical orbit utilized in  TDC II  is derived from  a 60-degree-of-freedom simulation for DFACS, performed by the Innovation Academy for Microsatellites of CAS~\cite{dfacs_simulation}. 
    The simulation is conducted with  quadruple-precision floating-point numbers, and incorporates the effects of 
    gravitational perturbations from celestial bodies, 
    the  forces acted by micro-thrusters, and solar radiation pressure, \emph{etc}. 
    As can be seen in FIG.~\ref{fig:LTTs_and_DPLs}, during the mission time of 1 year, the orbit has a maximum armlength difference of $\mathcal{O}(1\%)$ and maximum armlength variation rate of less than 5 m/s, fully compliant with Taiji's design requirements~\cite{taiji_2,taiji_orbit_simulation_requirement}. 

    \textbf{$\bullet$ Noises}: The noise data of TDC II originate from two distinct sources: those  related to the motions of SCs and TMs  are extracted from GRS \& DFACS simulation, including the ACC noise of TMs, the jitters of SCs, and the angular jitters of MOSAs (used to further simulate the inter-spacecraft TTL noise via a simple linear coupling model~\cite{ttl_noise_subtraction}), \emph{etc.}; while others are generated according to Taiji's design noise curves, \emph{e.g.} the readout noises of interferometers. 
    Additionally, glitches are simulated with the LISA Pathfinder legacy model~\cite{LPF_glitch,LDC_Spritz_manual}. 

\subsection{The TDC II Datasets}\label{subsec:4.2}

\begin{figure*}
    \centering
    \begin{minipage}[t]{0.32\textwidth}
        \includegraphics[width=\textwidth]{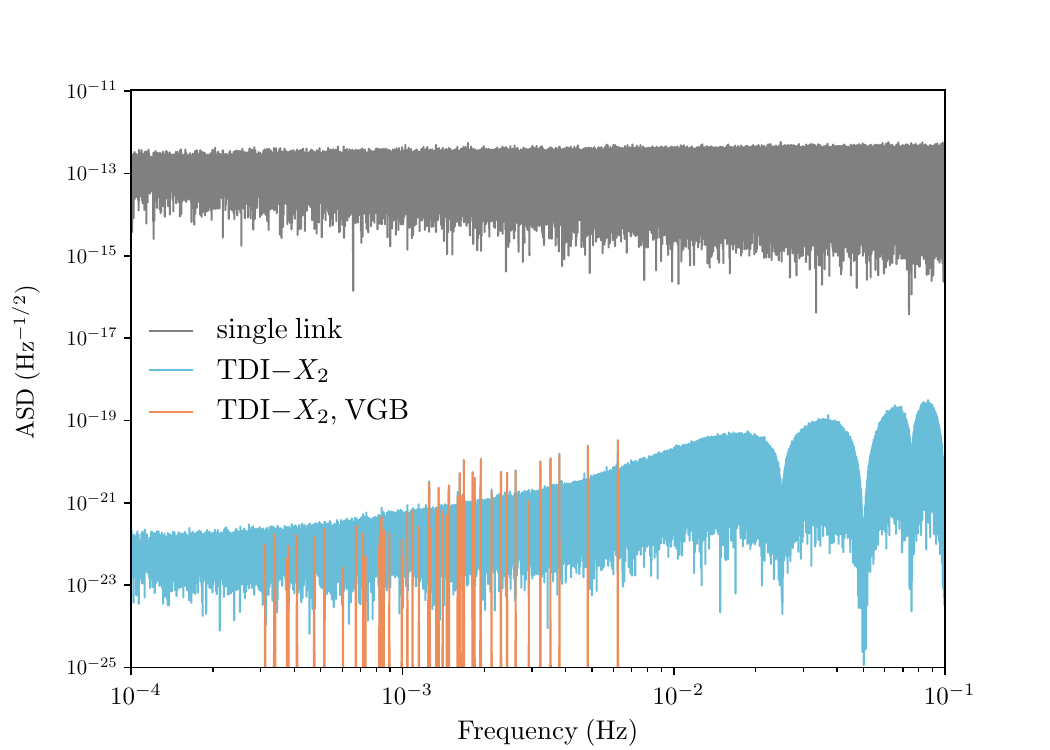}
        \subcaption{TDC II-0.1: VGBs (ASD).}
    \end{minipage}
    \hfill 
    \begin{minipage}[t]{0.32\textwidth}
        \includegraphics[width=\textwidth]{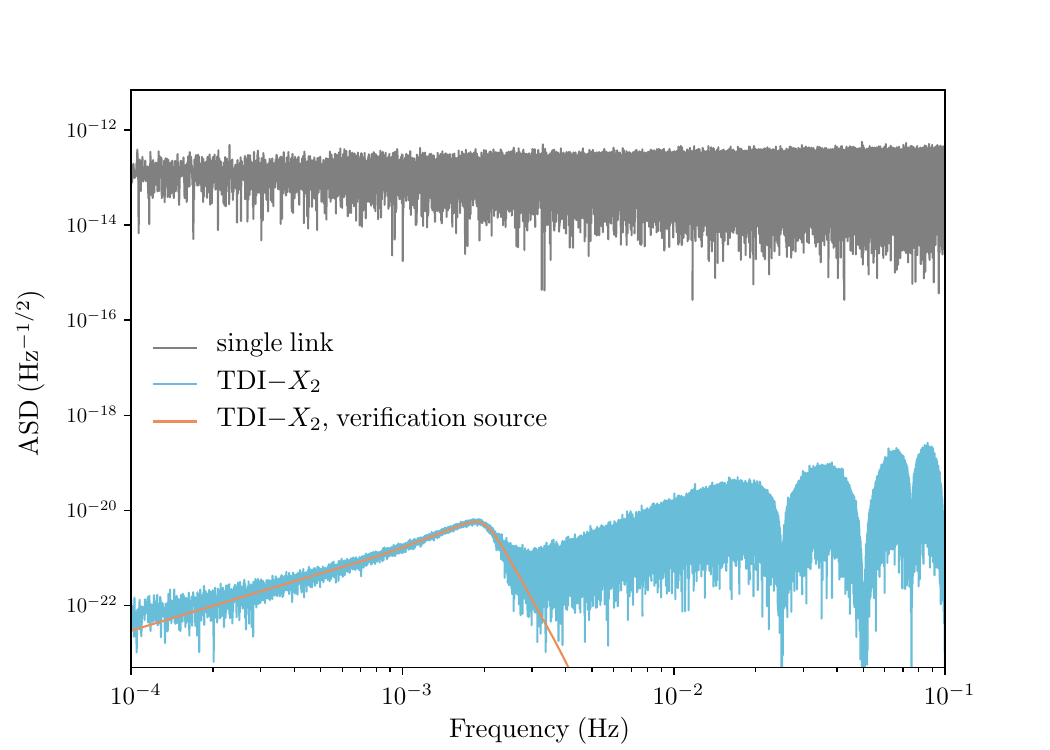}
        \subcaption{TDC II-0.2: single MBHB (ASD).}
    \end{minipage}
    \hfill
    \begin{minipage}[t]{0.32\textwidth}
        \includegraphics[width=\textwidth]{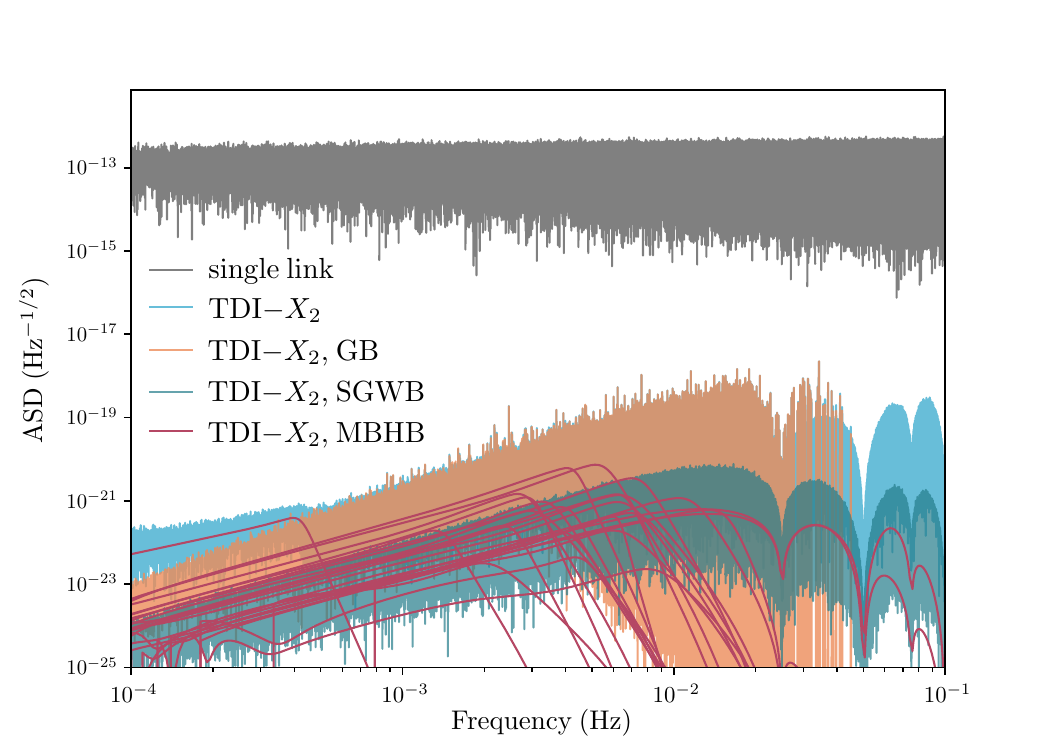}
        \subcaption{TDC II-1.1: multi-source (ASD).}
    \end{minipage}
    
    \vspace{0.2cm} 
    \begin{minipage}[t]{0.32\textwidth}
        \includegraphics[width=\textwidth]{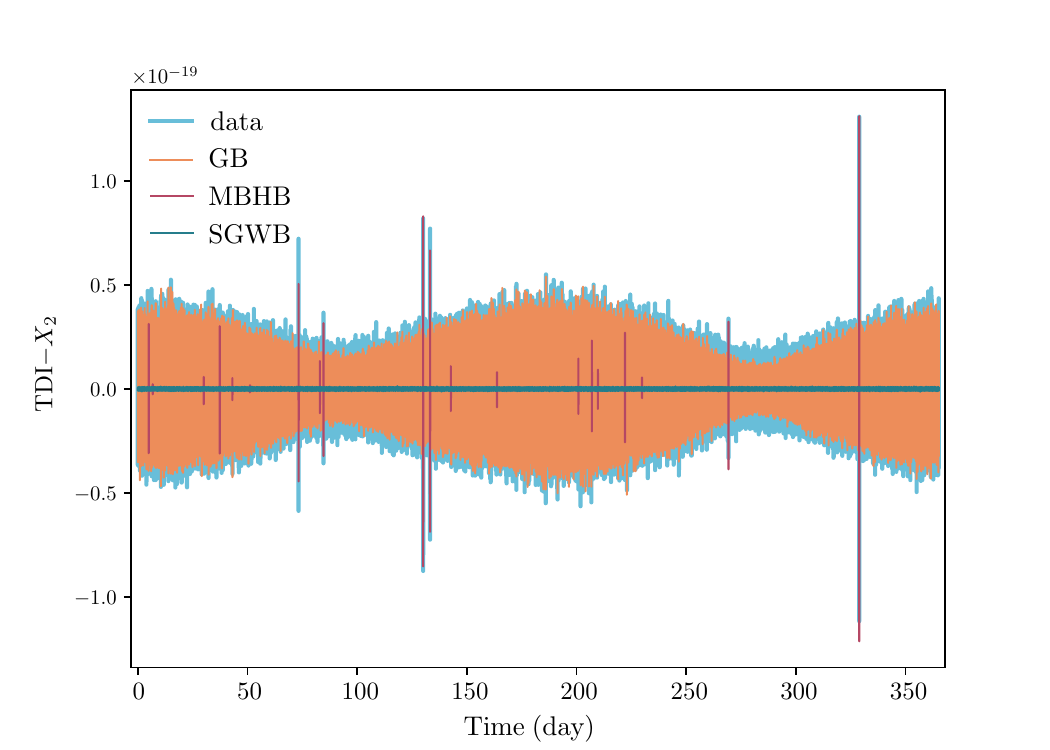}  
        \subcaption{TDC II-1.1: multi-source (time series, downsampled to 0.1 Hz).}
    \end{minipage}
    \hfill
    \begin{minipage}[t]{0.32\textwidth}
        \includegraphics[width=\textwidth]{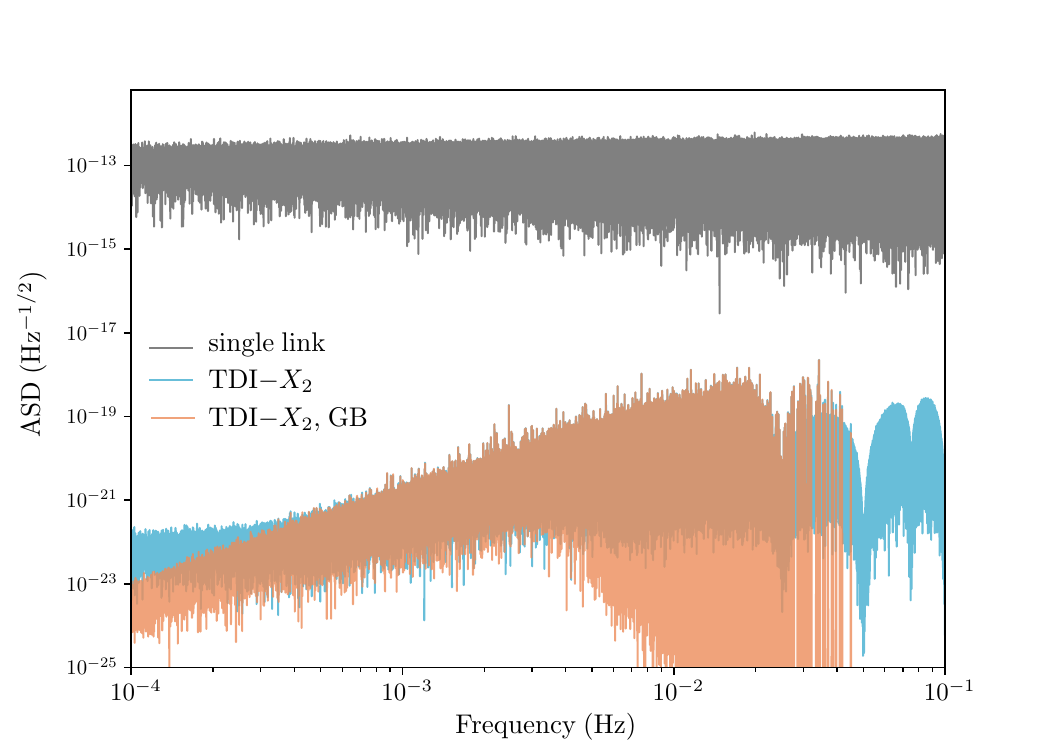}
        \subcaption{TDC II-2.1: GBs (ASD).}
    \end{minipage}
    \hfill
    \begin{minipage}[t]{0.32\textwidth}
        \includegraphics[width=\textwidth]{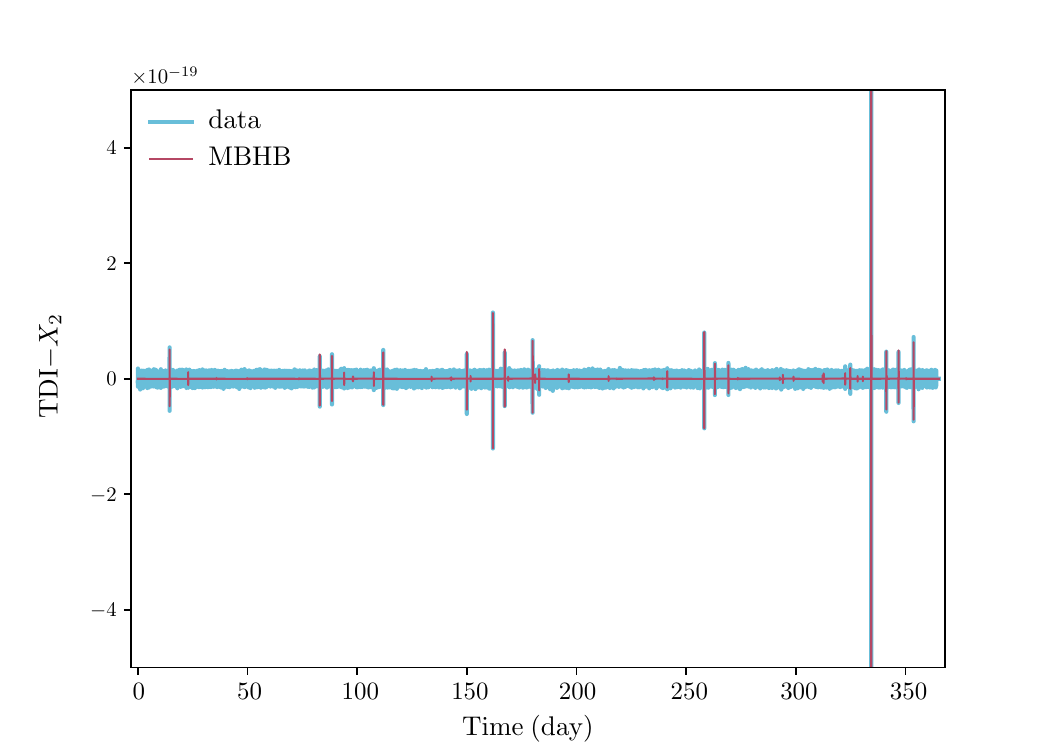}
        \subcaption{TDC II-2.2: overlapping MBHBs (time-series, downsampled to 0.1 Hz).}
    \end{minipage}
    
    \vspace{0.2cm}
    \begin{minipage}[t]{0.31\textwidth}
        \includegraphics[width=\textwidth]{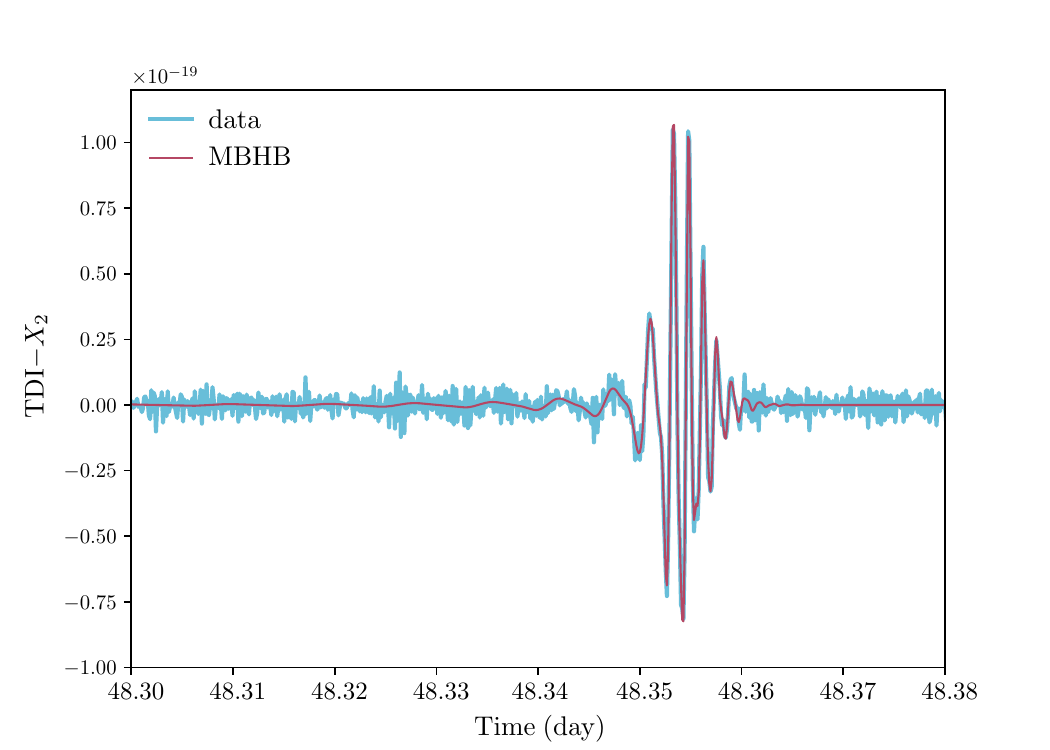}
        \subcaption{TDC II-2.3: MBHBs with eccentricity and higher harmonics (time-series, downsampled to 0.1 Hz, zoomed in around one of the mergers).}
    \end{minipage}
    \hfill
    \begin{minipage}[t]{0.32\textwidth}
        \includegraphics[width=\textwidth]{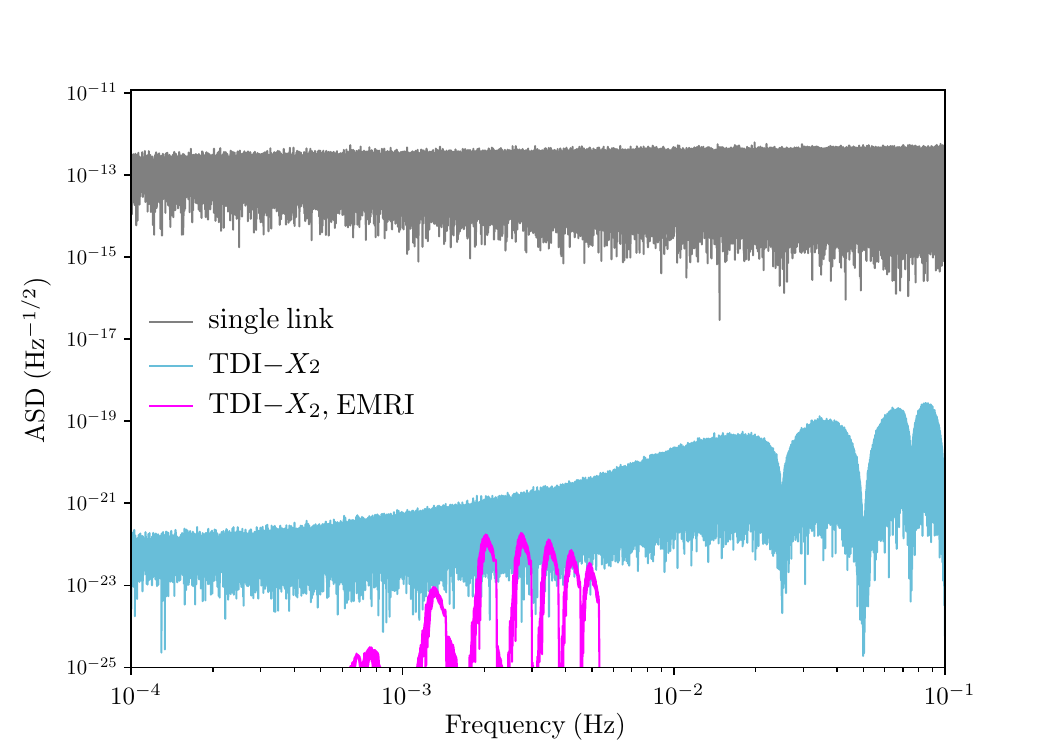}
        \subcaption{TDC II-2.5: EMRI (ASD, one of the four EMRIs is shown as a representative case).}
    \end{minipage}
    \hfill
    \begin{minipage}[t]{0.32\textwidth}
        \includegraphics[width=\textwidth]{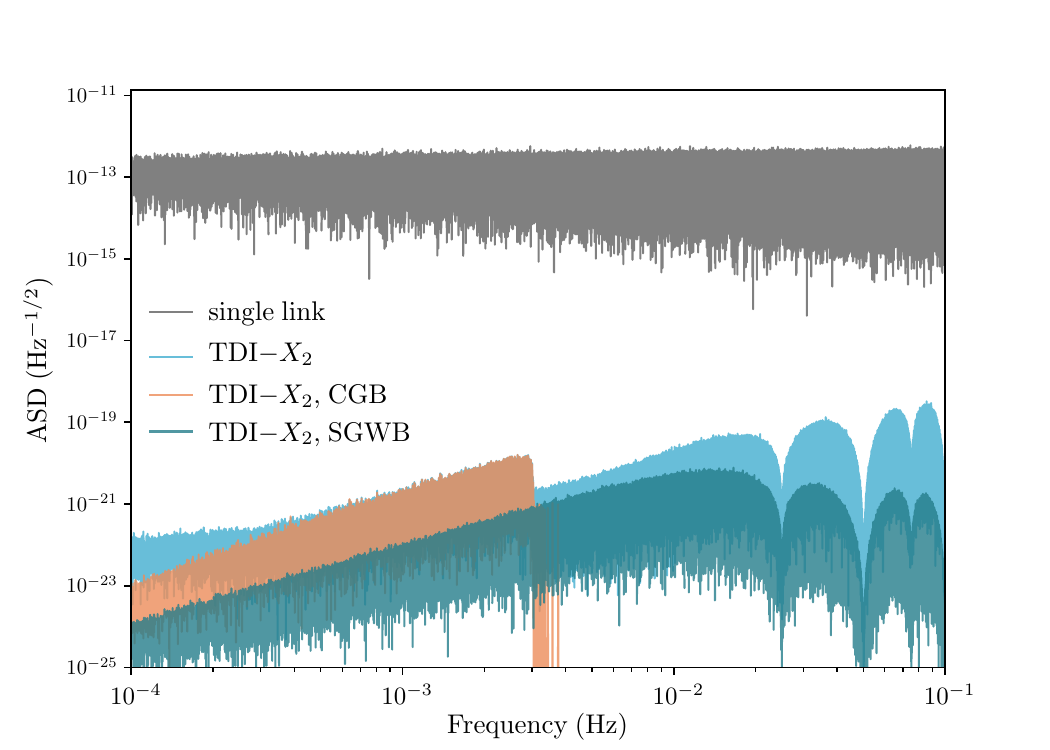}  
        \subcaption{TDC II-2.8: SGWB (ASD).}
    \end{minipage}
    
    \vspace{0.2cm}
    \begin{minipage}[t]{0.32\textwidth}
        \includegraphics[width=\textwidth]{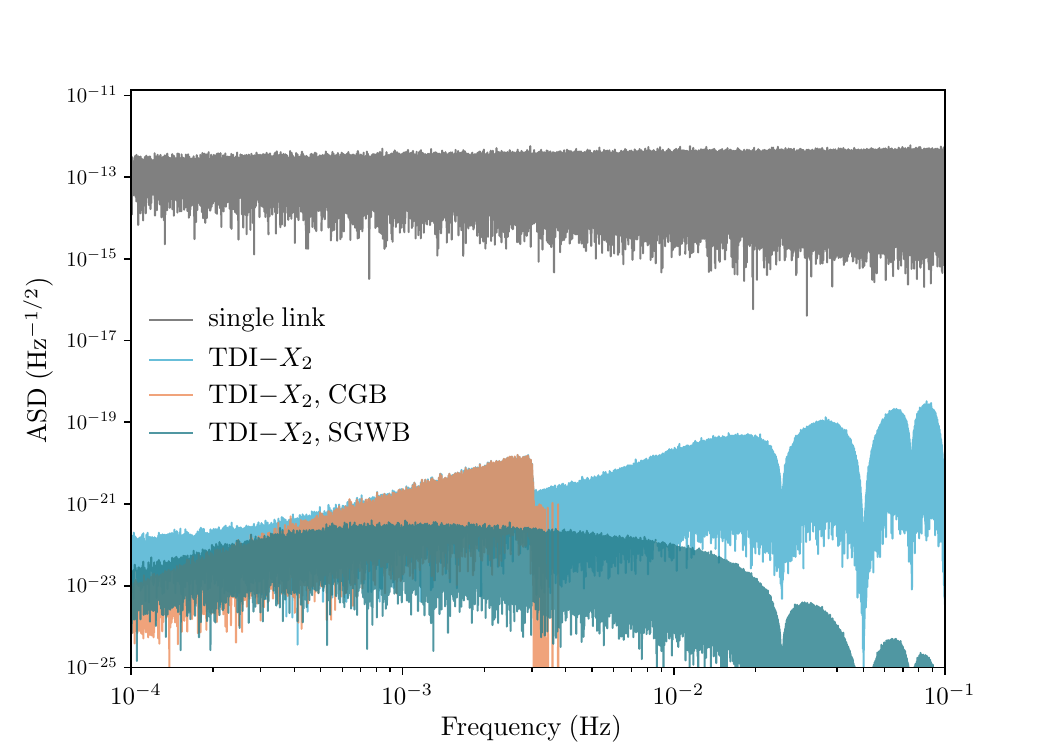}
        \subcaption{TDC II-2.9: SGWB (ASD, another profile).}
    \end{minipage}
    \hfill
    \begin{minipage}[t]{0.32\textwidth}
        \includegraphics[width=\textwidth]{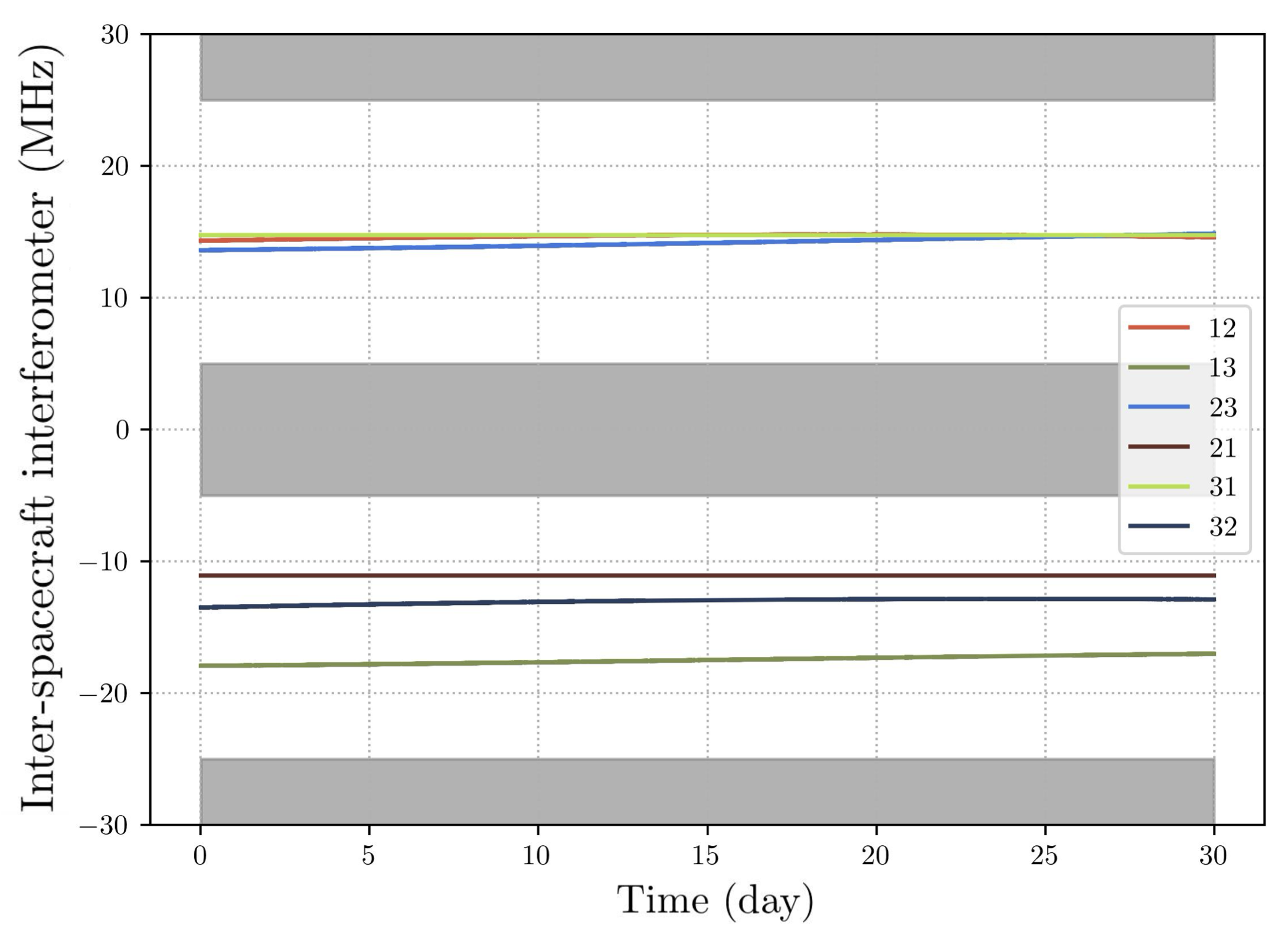}
        \subcaption{TDC II-3.1: raw ISI readouts (time series).}
    \end{minipage}
    \hfill
    \begin{minipage}[t]{0.32\textwidth}
        \includegraphics[width=\textwidth]{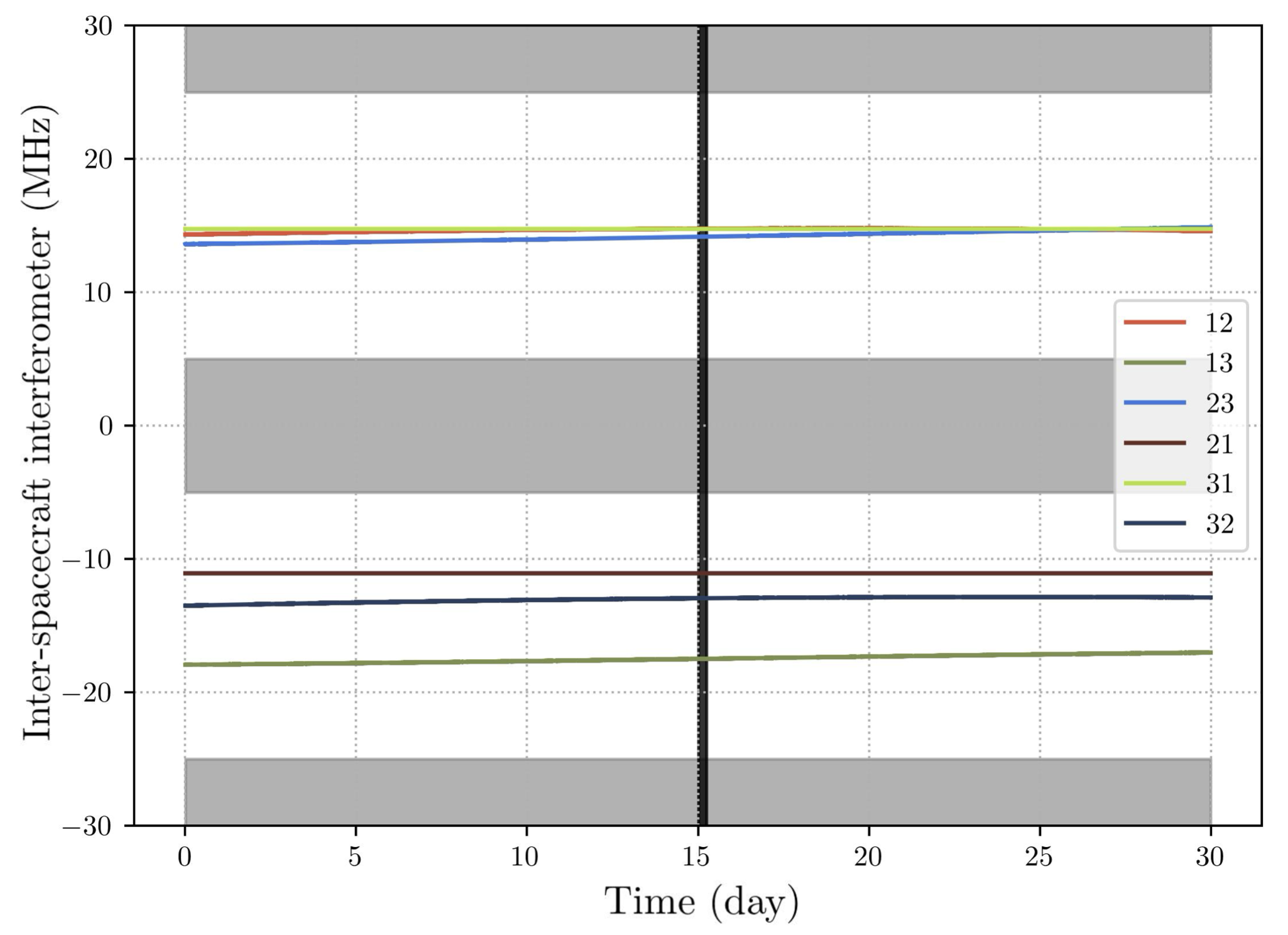}
        \subcaption{TDC II-4.1: raw ISI readouts (time series, gap shown as a black vertical line).}
    \end{minipage}
    
    \caption{The TDC II datasets,   either shown in the time domain or frequency domain (as ASDs) for optimal visualization.}
    \label{fig:tdc_data}
\end{figure*}


The current release of TDC II  comprises 5 groups of datasets (FIG.~\ref{fig:tdc_data}), each targeting specific challenge topics listed in Section~\ref{sec:3}.
Except for group 0 designed as illustrative examples for the access and use of TDC II, most of the remaining datasets are  \textbf{blind} (\emph{i.e.} the source parameters are unknown). 

Groups 1 and 2 focus on building the tools for scientific analysis. 
These data are presented in the form of single-link measurements $\eta_{ij}$ (in the fractional frequency difference unit, see Eq.~(\ref{eq:eta}) for  definition), instead of the $\{X, Y, Z\}$ or $\{A, E, T\}$ TDI combinations. 
One may notice that this is a significant change relative to the previous data challenges~\cite{MLDC,LDC,TDCI}, 
motivated by the need to systematically investigate the performances of diverse TDI configurations on GW detection, noise suppression and characterization, \emph{etc} (see Sections~\ref{subsec:3.1}, \ref{subsec:3.2}). 
Users may construct  customized  TDI variables utilizing the tools and examples provided by {the \texttt{Triangle} code.  } 
The injected laser frequency noise offers  an intrinsic  metric to validate the basic functionality of TDI schemes. 
Besides,  the  sampling rate  is set to 1 Hz (although the raw sampling rate of  telemetry data is  higher than 1 Hz, it should suffices for GW signals), and  all the data have been perfectly detrended and  synchronized to TCB. 
The injected noises types include laser frequency noise, TM ACC  noise, readout noises of interferometers (ISIs, RFIs, TMIs), optical path noises and fibre back link noise.

Groups 3 and 4 are designed to establish an end-to-end data analysis pipeline, 
therefore these two sets are presented in the form of raw laser interferometric measurements in the frequency (Hz) unit, namely the readouts of ISIs, RFIs, and TMIs, each including a carrier and a sideband measurement, 
together with the auxiliary data (pseudo ranges, angular jitters read by the differential wavefront sensors, orbit determination, measured clock deviations from TCB).
These data are recorded according to the on-board clocks, which deviate from TCB due to instrumental imperfections and relativistic corrections. 
The sampling rate is set to 4 Hz. 
Beyond the noises components  in groups 1 and 2, also incorporated in groups 3 and 4 are the clock noise, sideband modulation noise, SC jitter, TTL noise, as well as the readout noises of auxiliary data.

{
Despite the anticipated mission lifetime of at least five years, the maximum duration of the mock data is one year. 
This is primarily limited by the computational cost of spacecraft dynamics simulation. 
Due to the need to solve high-dimensional differential equations with quadruple-precision numbers, the generation of one-year orbital data typically takes several weeks. 
However, considering the realistic  scenario of future detection, we believe that the current duration of mock data remains viable for  algorithm development and  GW scientific research.  
For future space-based detection, to ensure the timeliness of scientific outputs,  global analysis may be conducted weekly or monthly as new data arrive, rather than after the whole mission lifetime. 
Therefore, simulated data spanning up to five years or longer is not always necessary.
Additionally, we also plan to incorporate longer-duration mock data into TDC II in subsequent updates.
}

Each dataset is stored in a separate HDF5 (\texttt{.h5}) file, whose download link can be found in the  website of TDC II~\cite{TDCII_website}. 
Within the HDF5 files, all the data are organized under the attribute ``\texttt{eta}'' (for groups 0, 1 and 2) or attributes ``\texttt{interspacecraft} \texttt{\_interferometer\_carrier}'', \emph{etc} (for groups 3 and 4), with the corresponding sampling times stored under ``\texttt{time}''. 
Except for these common features, in the following we present a detailed description on each of the dataset of TDC II:

    \textbf{0. The ``verification'' datasets}: In dataset 0.0 / 0.1, we inject 55 VGBs / 1 MBHB (IMRPhenomD waveform) upon the instrumental noises, respectively.
    Designed as simple datasets for illustration and validation,  all the source parameters can be found in the data files.
    The accompanying tutorials in the \texttt{Triangle-GB} and \texttt{Triangle-BBH} codes will demonstrate a simple workflow from loading data, TDI combination, waveform modeling, to full Bayesian posterior estimation.
    
    \textbf{1. The ``global fit'' dataset}: The multi-source  dataset of TDC II includes $\sim 4.5 \times 10^7$ GBs, 25 MBHBs, SGWB and instrumental noises. 
    Leveraging the well validated signal models in the former datasets, users may concentrate on exploring and developing  global analysis technique for realistic scenarios: numerical orbit, diverse (2nd-generation) TDI configurations, unknown spectral shapes of noise and SGWB, as well as a higher level of  signal overlapping (compared to LDC 2a and TDC I), \emph{etc}. 
    
    \textbf{2. The ``single source'' datasets}: Datasets in this group are tailored to address the analysis of specific signal types, including:\\ 
    \textbf{(1)} The global analysis of  GBs (dataset 2.1);\\ 
    \textbf{(2)} The  estimation of more overlapping MBHBs (dataset 2.2, 50 MBHBs / year, the IMRPhenomT waveform is utilized since it might be  suitable for time-domain analysis~\cite{phenomt1}) or individual MBHBs with complex waveforms (dataset 2.3, 3 MBHBs in total, SEOBNRv5EHM waveform);\\ 
    \textbf{(3)} The estimation of EMRI with AK, Kerr-BH, Bumpy-BH and b-EMRI waveforms (datasets 2.4, 2.5, 2.6, 2.7, notice that considering the widely recognized difficulty in EMRI estimation, the prior range of 2.4, as well as the parameters of more complex 2.5, 2.6, 2.7 will be provided in the data files, in order to support targeted search);\\ 
    \textbf{(4)}  the detection and discrimination of SGWBs  from  confusion foreground and instrumental noises (datasets 2.7, 2.8).
    
    \textbf{3. The ``end-to-end'' dataset}: 
    This dataset differs from the former ones by providing raw interferometric measurements (ISIs, TMIs, RFIs) alongside  auxiliary data, with extended noise types and the effect of onboard clock drift incorporated. 
    The TCB timestamps of data and the  delay times used in TDI combination are not  directly provided but must be derived through pre-processing (clock synchronization and inter-spacecraft ranging) instead. 
    After these, the noise floor of data can be further suppressed through the subtraction of inter-spacecraft TTL noise, which is generated with a simple linear coupling model~\cite{ttl_noise_subtraction}. 
    The injected GW signals comprise a few number of MBHBs and GBs. 
    This dataset aims to assist in developing  an end-to-end pipeline from raw data to scientific products, and  addressing the  coupling issues between pre-processing and scientific analysis (\emph{e.g.} TTL coefficient  calibration, TDI ranging).
    
    \textbf{4. The ``up-to-date'' dataset}: 
    To demonstrate the Taiji simulation group's progress in data simulation and illustrate  up-to-date knowledge on realistic data, a rolling updating dataset is also created. 
    In its initial release, this dataset is built upon the settings of ``end-to-end'' dataset, with more GW signals injected,  and   data anomalies such as glitches and gaps are also incorporated.

{It should be emphasized that the current version of TDC II  should  be regarded as  \textbf{representative}, rather than an exhaustive collection of  all the scientific objectives and challenging issues.} 
Users are encouraged to go beyond the default TDC II datasets by injecting custom signals and noises  using the  \texttt{Triangle} toolkit, thereby broadening the scope of exploration.
{We also plan to continuously update TDC II in the future (\emph{e.g.} by extending the duration of mock data, and  injecting  sBHB signals, which represent an important target for Taiji that was not included in the initial release). }

{
The timeline for TDC II is as follows.  
The official release is scheduled for May, 28, 2025. 
Prior to the official release, a beta test release will be deployed on May, 23, 2025. 
During this period, users are encouraged to propose suggestions regarding the optimization of datasets. 
The first collection of TDC II results is planned on October 9, 2025, and
the collected results will be summarized and made public on the website. 
The first dataset update will follow thereafter.}

\section{The TDC II toolkit: {\texttt{Triangle}}}\label{sec:5}
The TDC II toolkit, consisting of three code repositories prefixed with ``\texttt{Triangle}'', {is also released} alongside the mock data. 
Within this suite, \texttt{Triangle-Simulator} is a time-domain prototype  simulator for the data of space-based GW detectors.  
Given the models in Section~\ref{subsec:4.1} and under the settings of Section~\ref{subsec:4.2},  this simulator enables full reproducibility of the TDC II data.
It is designed to assist users in understanding the characteristics of Taiji data, 
creating their own mock data by injecting customized 
 noises and GW waveforms, hence 
uncovering new data analysis challenges and scientific objectives.
Furthermore, the models and functions within ~\texttt{Triangle-Simulator}, especially  the ones related to GW responses, provide benchmarks for developing and validating data analysis tools.
This simulator encapsulates the simulation of  GW responses, noises, instrumental effects (\emph{e.g.} clock deviations), TDI and other pre-processing steps in a unified pipeline. 
Based on a  modular design,  it supports  user-defined  models (or data) for  orbits, noises and signals. 
In principle \texttt{Triangle-Simulator} is generically applicable to any space-based GW detection mission with a triangular configuration, not specifically Taiji, but  also missions such as LISA and {TianQin}. 
Beyond the source codes, the repository also offers 5  tutorials, {covering the  introductions on}  laser interferometric measurements, noise types and transfer functions, GW  response formalism, and TDI combinations, followed by basic instructions on the access and usage of TDC II data. 

The other two supplementary repositories are designed to provide illustrative examples for users (particularly students), showcasing how to perform  simple Bayesian analysis on some easy tasks. 
They are mainly based on two 
frequency-domain fast signal response calculators: \texttt{Triangle-BBH} targeting MBHBs  and \texttt{Triangle-GB} specialized for GBs. 
Under identical configurations (orbit, waveform approximants, source parameters, \emph{etc}), 
the  frequency-domain responses are consistent with 
the time-domain simulations of \texttt{Triangle-Simulator}.
More specifically, \texttt{Triangle-BBH} models the 2nd-generation TDI responses of MBHBs in the frequency domain, utilizing  two implementations of the IMRPhenomD/HM approximants: one is CPU-based \texttt{WF4PY}~\cite{wf4py_github,wf4py1,wf4py2}, and the other is GPU-accelerated \texttt{BBHx}~\cite{bbhx_zenodo,bbhx1,bbhx2}. 
The interface for numerical orbit is the same as \texttt{Triangle-Simulator}. 
The speed of \texttt{Triangle-BBH} is generally 0.1 ms - 1 ms per waveform evaluation (depending on the length of frequency series and the capability of hardware), which enables Bayesian parameter estimation of MBHBs. 
On the other hand, \texttt{Triangle-GB} calculates the 2nd-generation TDI responses of GBs, which is adapted  from \texttt{GBGPU}~\cite{gbgpu_zenodo,gbgpu1,gbgpu2} (the GPU implementation of \texttt{FastGB} algorithm~\cite{fastgb}) in order to support the numerical orbit of Taiji and 2nd-generation TDI combinations. 
The speed is up to 10 $\mu$s to 100 $\mu$s per GB waveform. 
Examples for the analysis of TDC ``verification'' datasets 0.1 and 0.2 are offered in these repositories, with one of them based on \texttt{Triangle-GB} shown in  Appendix~\ref{sec:Appendix}.
One should notice that these examples should only  be seen as   illustrative, but not solutions to all the challenges.

  

\section{Summary and Outlook}\label{sec:6}



{In this paper, we systematically reviewed   the anticipated 
 data analysis challenges  for the realistic observation of  Taiji}, and introduced a new simulation testbed for data analysis, including 
 TDC II, a collection of mock datasets intended to support the research towards addressing these challenges, and an open-source toolkit \texttt{Triangle}.
 Collectively, they will assist in  the development of Taiji's data analysis tools in the upcoming stage, 
 
In the future, the Taiji simulation group will continue enhancing the realism and reliability of mock data  by incorporating  
more physical models for the GRSs, the far-field optics (especially the wavefront errors caused by telescope aberrations and their far-field propagation)~\cite{ttl_model}, the impacts of thermal stability  on core metrology systems, the effects of stray lights~\cite{Sasso_2019}, 
the onboard optical delays~\cite{tdi_on_board_delay}, the effects of digitization and downsampling filters~\cite{tdi_filter,tdi_aa_filter}, \emph{etc},  and further integrate the readouts of auxiliary sensors to support the  development  of more ``end-to-end'' pipelines.
Especially, keen attentions will be paid to the models  that have been validated via   ground-based experiments. 
Besides,  the GW source catalogs and waveforms will also be continuously updated, 
in alignment with the frontier developments in astronomy and cosmology.

Looking ahead, joint space-based detection with future LISA-Taiji or LISA-Taiji-TianQin networks~\cite{lisa_taiji_network1,lisa_taiji_network2,lisa_taiji_tianqin_network1,lisa_taiji_tianqin_network2} could enhance the detectability of  GW signals, hence  enabling  unprecedented tests of  astrophysics, cosmology and fundamental physics. 
Therefore, future TDC should also expand its  scope to include  other detectors.
Researchers would  have the opportunity to   leverage cross-mission data to refine noise calibration and mitigation, improve source localization, and most importantly, perform high-confidence detection of SGWBs.

\begin{acknowledgments}
The development of TDC II is supported by National Key Research and Development Program of China (Grant No. 2024YFC2207300, No. 2021YFC2201903, No. 2021YFC2201901, No. 2020YFC2200100). 
The creation  of mock data significantly depends on the DFACS simulation conducted by the Innovation Academy for Microsatellites of CAS. 
We acknowledge the contributions of National Space Science Center of CAS, Shanghai Astronomical Observatory of CAS, and Peking University for providing essential noise models, GW waveform data and GW source catalogs. 
Collaborative support was also offered by the University of Chinese Academy of Sciences, Beijing Normal University, as well as other institutes, groups and individuals. 
{Also, we are grateful for the anonymous reviewers’ comments and suggestions during the review process. }
For the use of  \texttt{Triangle} in your published research works, please cite the references provided in the README files. 
\end{acknowledgments}

\bibliographystyle{apsrev4-2}
\bibliography{apssamp}

\begin{figure*}
    \centering
    \includegraphics[width=0.85\linewidth]{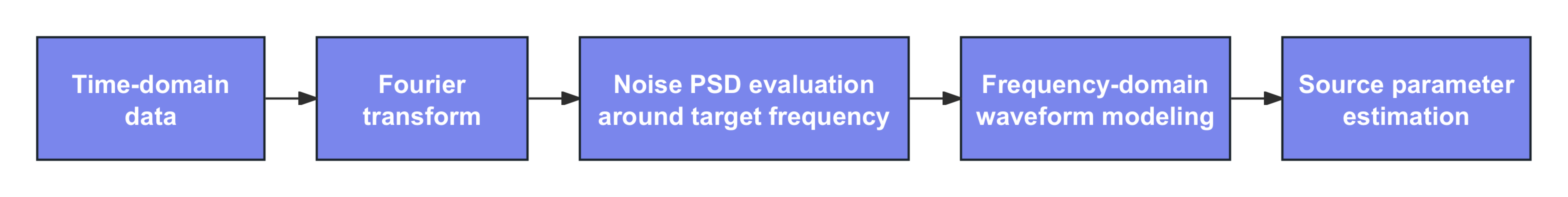}
    \caption{The ``noise-agnostic'' workflow implemented on the verification binary dataset.}
    \label{fig:noise_agnostic_workflow}
\end{figure*}

\begin{figure*}
    \centering
    \includegraphics[width=\linewidth]{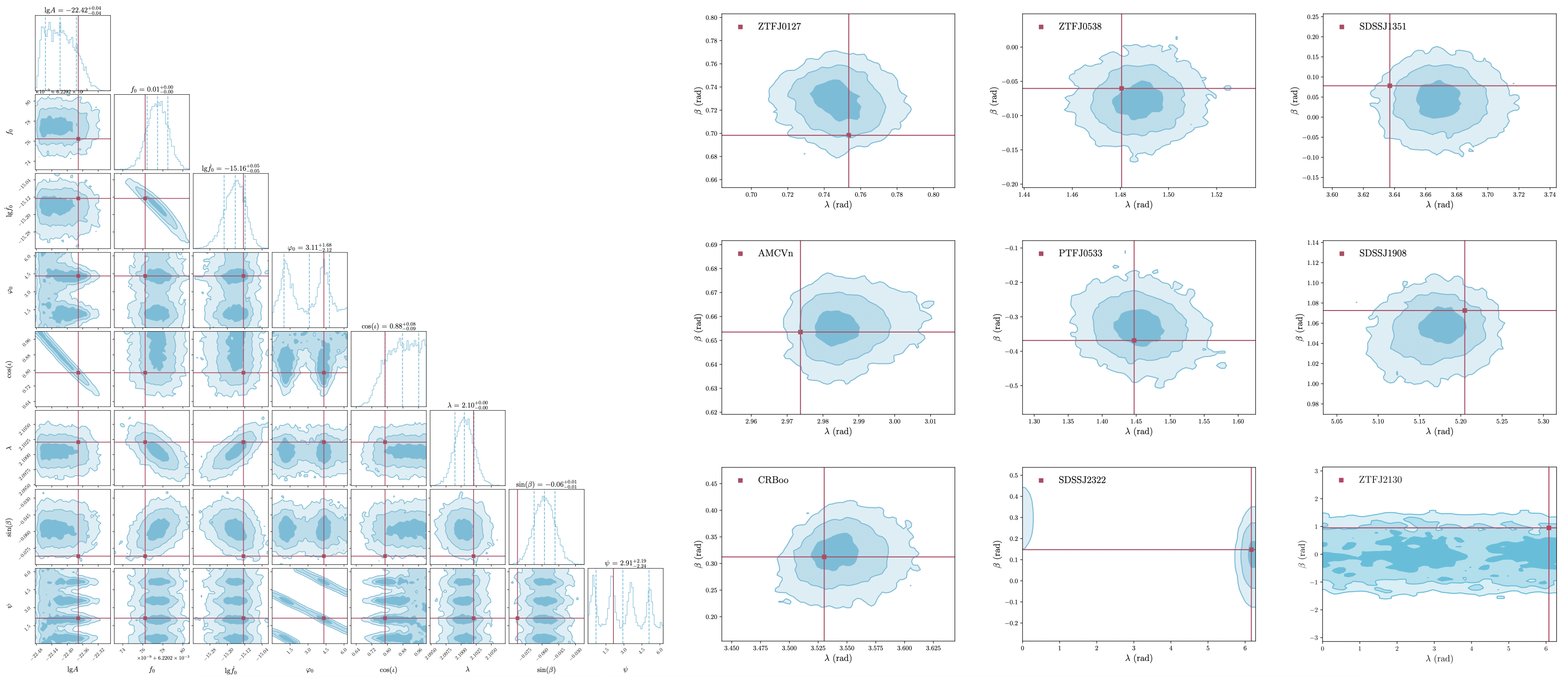}
    \caption{The full posterior distributions for 1 VGB and the sky localizations of other  9  VGBs.}
    \label{fig:VGB_posterior}
\end{figure*}

\appendix
\section{Parameter estimation of verification binary dataset with a ``noise agnostic'' workflow}\label{sec:Appendix}

In this appendix, we showcase a simple parameter estimation task on dataset 0.1, 
illustrating the basic utility of TDC II data and validating the fidelity of our simulations. 
The data is composed  of  55 VGB signals  and instrumental noises. 
As discussed in Section~\ref{subsec:3.2}, we are very likely to face a situation where noise characteristics can not be known priorly.
To account for this scenario, our analysis is conducted based on  a ``noise-agnostic'' workflow (FIG.~\ref{fig:noise_agnostic_workflow}). 
After suppressing laser frequency noises via the  2nd-generation Michelson TDI combinations,  we analyze  a narrow frequency bin of width $\mathcal{O}(10^{-6})$ Hz around each signal, and utilize data at the adjacent frequency bands to evaluate the  PSD via the Welch method.  
The mean PSD of these bands is adopted as the noise floor (ignoring the spectral shape due to the narrow bandwidth).
Now we are ready to compute the likelihood function using \texttt{Triangle-GB} and generate posterior samples with  PTMCMC implemented in the \texttt{Eryn}~\cite{eryn,eryn_zenodo,eryn_emcee} sampler.
Conservative hyperparameters (\texttt{n\_walker}=200, \texttt{n\_temp}=10)  are chosen, and an affine invariant  proposal is used  to minimize tuning.

The full posterior distributions for 1 VGB and the sky localizations of other  9 randomly selected VGBs are shown in FIG.~\ref{fig:VGB_posterior}.
Note that  our simulation  spans only one year, while the  ``detectable'' GBs~\cite{VGB_paper} are defined based on 48-month observation. 
Consequently, the last signal exhibits a low total SNR ($\lesssim$ 7), which is faithfully reflected in its posterior distribution (\emph{i.e.} the localization is poorly constrained).

\end{document}